# Compounding Fields and Their Quantum Equations in the Curved Spaces

Zihua Weng

*School of Physics and Mechanical & Electrical Engineering, P. O. Box 310,
Xiamen University, Xiamen 361005, China*

**Abstract**

Based on the concept of curved spacetime in Einsteinian General Relativity, the field theories and their quantum theories in the curved octonion spaces etc are discussed. The research results discover the close relationships of the curved spacetimes with the field theories and quantum theories. In the field theories of curved spacetimes, the curvatures have direct effect on field strength and field source etc. While the curvatures have direct impact on wave functions and quantum equations etc in the quantum theories of curved spacetimes. The research results discover that some abnormal phenomena of field source particles could be explained by the field theories or quantum theories in the curved spacetimes.

## 1. Introduction

The viewpoint of curved spacetime should be expanded from the curved Four-spacetime to the curved quaternion spacetime etc. The paper describes the field theories and their quantum theories in the curved quaternion, octonion, sedenion and trigintaduonion spacetimes, and then draws some conclusions which are consistent with the Einsteinian General Relativity theory, Maxwellian electromagnetic field theory and the Newtonian gravitational field theory etc. A few prediction associated with movement feature of different kinds of matter can be deduced, and some new and unknown experiment phenomena can be expected.

## 2. Electromagnetic field in curved quaternion space

The electromagnetic field theory can be described with the straight quaternion space. In the paper, the spacetime should be expanded from the straight spacetime to the curved spacetime. The curved spacetime, which is associated with the electromagnetic interaction and possesses the physics content, is adopted by the curved quaternion space.

---

*E-mail Addresses*: xmuwzh@hotmail.com, xmuwzh@xmu.edu.cn

Table 1. The quaternion multiplication table

|   | 1 | $\vec{i}$ | $\vec{j}$ | $\vec{k}$ |
|---|---|---|---|---|
| 1 | 1 | $\vec{i}$ | $\vec{j}$ | $\vec{k}$ |
| $\vec{i}$ | $\vec{i}$ | -1 | $\vec{k}$ | $-\vec{j}$ |
| $\vec{j}$ | $\vec{j}$ | $-\vec{k}$ | -1 | $\vec{i}$ |
| $\vec{k}$ | $\vec{k}$ | $\vec{j}$ | $-\vec{i}$ | -1 |

In the local curved quaternion space (E space, for short) of electromagnetic interaction, the base $\mathcal{E}_e = (1, \vec{i}, \vec{j}, \vec{k})$, and the displacement $\mathcal{R}_e = (r^0, r^1, r^2, r^3)$. Where, $r^0 = ct$, c is the speed of light beam, and t denotes the time.

In curved quaternion space, the differential operator should be expanded from $\partial_i$ to $\nabla_i$.

$$\nabla_i A^k = \partial_i A^k + \Sigma \Gamma_{ji}{}^k A^j \tag{1}$$

wherein, $\Gamma_{ji}{}^k$ is a coefficient; $A^k$ is a physical quantity; $\partial_i = \partial / \partial r^i$; i, j, k = 0, 1, 2, 3.

The differential operator $\diamondsuit_e$ and its conjugate operator $\diamondsuit^*_e$ in the curved quaternion space are defined as

$$\diamondsuit_e = \nabla_0 + \vec{i}\,\nabla_1 + \vec{j}\,\nabla_2 + \vec{k}\,\nabla_3 \quad, \quad \diamondsuit^*_e = \nabla_0 - \vec{i}\,\nabla_1 - \vec{j}\,\nabla_2 - \vec{k}\,\nabla_3 \tag{2}$$

The electromagnetic field potential $\mathcal{A}_e(a^0, a^1, a^2, a^3)$ is defined as

$$\mathcal{A}_e = \diamondsuit^*_e \circ \mathcal{X}_e = (a^0, a^1, a^2, a^3) \tag{3}$$

where, $\mathcal{X}_e$ is the physical quantity in E space.

2.1 Field strength

The electromagnetic field strength $\mathcal{B}_e(B^0, B^1, B^2, B^3)$ is defined as

$$\begin{aligned}
\mathcal{B}_e &= \diamondsuit_e \circ \mathcal{A}_e \\
&= b^0 + (\vec{i}\,b_0{}^1 + \vec{j}\,b_0{}^2 + \vec{k}\,b_0{}^3) + (\vec{i}\,b_3{}^2 + \vec{j}\,b_1{}^3 + \vec{k}\,b_2{}^1) \\
&= (\nabla_0 a^0 - \nabla_1 a^1 - \nabla_2 a^2 - \nabla_3 a^3) \\
&\quad + \{\vec{i}\,(\nabla_1 a^0 + \nabla_0 a^1) + \vec{j}\,(\nabla_2 a^0 + \nabla_0 a^2) + \vec{k}\,(\nabla_3 a^0 + \nabla_0 a^3)\} \\
&\quad + \{\vec{i}\,(\nabla_2 a^3 - \nabla_3 a^2) + \vec{j}\,(\nabla_3 a^1 - \nabla_1 a^3) + \vec{k}\,(\nabla_1 a^2 - \nabla_2 a^1)\} \\
&= (\partial_0 a^0 - \partial_1 a^1 - \partial_2 a^2 - \partial_3 a^3) \\
&\quad + \{\vec{i}\,(\partial_1 a^0 + \partial_0 a^1) + \vec{j}\,(\partial_2 a^0 + \partial_0 a^2) + \vec{k}\,(\partial_3 a^0 + \partial_0 a^3)\} \\
&\quad + \{\vec{i}\,(\partial_2 a^3 - \partial_3 a^2) + \vec{j}\,(\partial_3 a^1 - \partial_1 a^3) + \vec{k}\,(\partial_1 a^2 - \partial_2 a^1)\} \\
&\quad + \Sigma\,(\Gamma_{j0}{}^0 - \Gamma_{j1}{}^1 - \Gamma_{j2}{}^2 - \Gamma_{j3}{}^3)\,a^j \\
&\quad + \Sigma\{\vec{i}\,(\Gamma_{j1}{}^0 + \Gamma_{j0}{}^1) + \vec{j}\,(\Gamma_{j2}{}^0 + \Gamma_{j0}{}^2) + \vec{k}\,(\Gamma_{j3}{}^0 + \Gamma_{j0}{}^3)\}\,a^j \\
&\quad + \Sigma\{\vec{i}\,(\Gamma_{j2}{}^3 - \Gamma_{j3}{}^2) + \vec{j}\,(\Gamma_{j3}{}^1 - \Gamma_{j1}{}^3) + \vec{k}\,(\Gamma_{j1}{}^2 - \Gamma_{j2}{}^1)\}\,a^j
\end{aligned} \tag{4}$$

where, the sum of the first and fourth items in the right of equal mark is the gauge definition $b^0$; the sum of the second and fifth items is the electric field intensity ($b_0{}^1, b_0{}^2, b_0{}^3$); the sum of the third and sixth items is the flux density ($b_3{}^2, b_1{}^3, b_2{}^1$). Taking $b^0 = 0$, then the gauge condition of field potential can be achieved in the curved quaternion space.

2.2 Field source

The electromagnetic field source $\mathcal{S}_e(s^0, s^1, s^2, s^3)$ is defined as

$$\mu S_e = \Diamond^*_e \circ \mathcal{B}_e$$
$$= (\Diamond^*_e \circ \Diamond_e) \circ \mathcal{A}_e$$
$$= \Diamond_e^2 (a^0 + \vec{i} a^1 + \vec{j} a^2 + \vec{k} a^3) \tag{5}$$

where, $\Diamond_e^2 = \Diamond^*_e \circ \Diamond_e$. $\mu$ is a coefficient. Some identities equations of the field strength can be obtained from the above equation. Taking $S_e = (s^0 + \vec{i} s^1 + \vec{j} s^2 + \vec{k} s^3)$, then equations set of field potential in the electromagnetic field can be attained

$$\Diamond_e^2 (a^0 + \vec{i} a^1 + \vec{j} a^2 + \vec{k} a^3) = \mu (s^0 + \vec{i} s^1 + \vec{j} s^2 + \vec{k} s^3) \tag{6}$$

The operator $\Diamond_e^2$ in the curved quaternion space can be written as

$$\Diamond_e^2 = \Diamond^*_e \circ \Diamond_e$$
$$= (\nabla_0 - \vec{i} \nabla_1 - \vec{j} \nabla_2 - \vec{k} \nabla_3) \circ (\nabla_0 + \vec{i} \nabla_1 + \vec{j} \nabla_2 + \vec{k} \nabla_3)$$
$$= (\nabla_0 \nabla_0 + \nabla_1 \nabla_1 + \nabla_2 \nabla_2 + \nabla_3 \nabla_3)$$
$$+ \{\vec{i} (\nabla_0 \nabla_1 - \nabla_1 \nabla_0) + \vec{j} (\nabla_0 \nabla_2 - \nabla_2 \nabla_0) + \vec{k} (\nabla_0 \nabla_3 - \nabla_3 \nabla_0)\}$$
$$+ \{\vec{i} (\nabla_3 \nabla_2 - \nabla_2 \nabla_3) + \vec{j} (\nabla_1 \nabla_3 - \nabla_3 \nabla_1) + \vec{k} (\nabla_2 \nabla_1 - \nabla_1 \nabla_2)\}$$

In the above equation,

$$(\nabla_m \nabla_k - \nabla_k \nabla_m) a^i = \Sigma R_{kmp}{}^i a^p \tag{7}$$
$$R_{kmp}{}^i = \partial_m \Gamma_{pk}{}^i - \partial_k \Gamma_{pm}{}^i + \Sigma \Gamma_{qm}{}^i \Gamma_{pk}{}^q - \Sigma \Gamma_{qk}{}^i \Gamma_{pm}{}^q \tag{8}$$

where, $R_{kmp}{}^i$ is the curvature of curved quaternion space; m, n, p, q = 0, 1, 2, 3.

Then the definition of field source in the curved quaternion space can be rewritten as

$$\mu S_e = (\Diamond^*_e \circ \Diamond_e) \circ \mathcal{A}_e = \mu S_{eX} + \mu S_{eR}$$
$$= \{\vec{i} (\nabla_0 \nabla_1 - \nabla_1 \nabla_0) + \vec{j} (\nabla_0 \nabla_2 - \nabla_2 \nabla_0) + \vec{k} (\nabla_0 \nabla_3 - \nabla_3 \nabla_0)$$
$$+ \vec{i} (\nabla_3 \nabla_2 - \nabla_2 \nabla_3) + \vec{j} (\nabla_1 \nabla_3 - \nabla_3 \nabla_1) + \vec{k} (\nabla_2 \nabla_1 - \nabla_1 \nabla_2)$$
$$+ (\nabla_0 \nabla_0 + \nabla_1 \nabla_1 + \nabla_2 \nabla_2 + \nabla_3 \nabla_3)\} \circ (a^0 + \vec{i} a^1 + \vec{j} a^2 + \vec{k} a^3) \tag{9}$$

wherein,

$$\mu S_{eX} = (\nabla_0 \nabla_0 + \nabla_1 \nabla_1 + \nabla_2 \nabla_2 + \nabla_3 \nabla_3) \circ (a^0 + \vec{i} a^1 + \vec{j} a^2 + \vec{k} a^3)$$
$$\mu S_{eR} = \{\vec{i} (\nabla_0 \nabla_1 - \nabla_1 \nabla_0) + \vec{j} (\nabla_0 \nabla_2 - \nabla_2 \nabla_0) + \vec{k} (\nabla_0 \nabla_3 - \nabla_3 \nabla_0)$$
$$+ \vec{i} (\nabla_3 \nabla_2 - \nabla_2 \nabla_3) + \vec{j} (\nabla_1 \nabla_3 - \nabla_3 \nabla_1) + \vec{k} (\nabla_2 \nabla_1 - \nabla_1 \nabla_2)\}$$
$$\circ (a^0 + \vec{i} a^1 + \vec{j} a^2 + \vec{k} a^3)$$

If $\mu S_{eR} = 0$, the spacetime of the electromagnetic field is the straight quaternion space, and $\mu S_e = \mu S_{eX}$; If $\mu S_{eR} \neq 0$, the spacetime is the curved quaternion space, and $\mu S_e - \mu S_{eX} = \mu S_{eR}$; If $\mu S_{eR} \neq 0$ and $\mu S_e = 0$, the spacetime is the curved quaternion space, then $\mu S_{eX} = -\mu S_{eR}$. From the last equation, we can obtain the field equations $R_{kmp}{}^i = f(\mu S_{eX}, \mathcal{A}_e)$, which are similar to the field equations of Einsteinian General Relativity in the curved Four-spacetime.

In curved quaternion space, other physical quantities including the coefficient $\Gamma_{pk}{}^i$ and/or curvature $R_{kmp}{}^i$ can be obtained in the Table 2.

Table 2. The curvature of curved quaternion space in other physical quantities

|   | physical quantity | coefficient, $\Gamma_{pk}{}^i$ | curvature, $R_{kmp}{}^i$ |
|---|---|---|---|
| 1 | $\mathcal{X}_e$ | $\mathcal{A}_e = \Diamond^*_e \circ \mathcal{X}_e$ | $\mathcal{B}_e = \Diamond_e^2 \circ \mathcal{X}_e$ |
| 2 | $\mathcal{A}_e$ | $\mathcal{B}_e = \Diamond_e \circ \mathcal{A}_e$ | $\mu S_e = \Diamond_e^2 \circ \mathcal{A}_e$ |
| 3 | $\mathcal{B}_e$ | $\mu S_e = \Diamond^*_e \circ \mathcal{B}_e$ | $\mu J_e = \Diamond_e^2 \circ \mathcal{B}_e$ |
| 4 | $\mathcal{M}_e$ | $\mathcal{W}_e = c \Diamond^*_e \circ \mathcal{M}_e$ | $\mathcal{N}_e = c^2 \Diamond_e^2 \circ \mathcal{M}_e$ |

2.3 Variation rate of field source

The variation rate of electromagnetic field source $\mathcal{J}_e$ ($J^0$, $J^1$, $J^2$, $J^3$) in the curved quaternion space is defined as

$$\begin{aligned}\mathcal{J}_e &= \diamondsuit_e \circ \mathcal{S}_e \\ &= J^0 + (\vec{i}\, J_0^1 + \vec{j}\, J_0^2 + \vec{k}\, J_0^3) + (\vec{i}\, J_3^2 + \vec{j}\, J_1^3 + \vec{k}\, J_2^1) \\ &= (\partial_0 s^0 - \partial_1 s^1 - \partial_2 s^2 - \partial_3 s^3) \\ &\quad + \{\vec{i}\,(\partial_1 s^0 + \partial_0 s^1) + \vec{j}\,(\partial_2 s^0 + \partial_0 s^2) + \vec{k}\,(\partial_3 s^0 + \partial_0 s^3)\} \\ &\quad + \{\vec{i}\,(\partial_2 s^3 - \partial_3 s^2) + \vec{j}\,(\partial_3 s^1 - \partial_1 s^3) + \vec{k}\,(\partial_1 s^2 - \partial_2 s^1)\} \\ &\quad + \Sigma\, (\Gamma_{j0}^{\ 0} - \Gamma_{j1}^{\ 1} - \Gamma_{j2}^{\ 2} - \Gamma_{j3}^{\ 3})\, s^j \\ &\quad + \Sigma\{\vec{i}\,(\Gamma_{j1}^{\ 0} + \Gamma_{j0}^{\ 1}) + \vec{j}\,(\Gamma_{j2}^{\ 0} + \Gamma_{j0}^{\ 2}) + \vec{k}\,(\Gamma_{j3}^{\ 0} + \Gamma_{j0}^{\ 3})\}\, s^j \\ &\quad + \Sigma\{\vec{i}\,(\Gamma_{j2}^{\ 3} - \Gamma_{j3}^{\ 2}) + \vec{j}\,(\Gamma_{j3}^{\ 1} - \Gamma_{j1}^{\ 3}) + \vec{k}\,(\Gamma_{j1}^{\ 2} - \Gamma_{j2}^{\ 1})\}\, s^j \end{aligned} \quad (10)$$

Taking the gauge condition $J^0 = 0$, then the continuity equation of field source of the electromagnetic field can be achieved.

In the curved quaternion spacetime, the various properties among the field potential, field strength and field source of the electromagnetic interaction can be described by the curved quaternion space theory. According to the 'SpaceTime Equality Postulation' [1, 2], it can be deduced that spacetime derived from gravitational interaction is supposed to be the curved quaternion space also. Given that the coupling influence of fundamental interaction is neglected, the gravitational interaction is similar to the electromagnetic interaction and can also be described by the curved quaternion space.

## 3. Electromagnetic-gravitational field in curved octonion space

The electromagnetic interaction and gravitational interaction can be both described in the curved quaternion space. By means of the conception of space expansion etc, two types of curved quaternion spaces can combine into a curved octonion space. In curved octonion space, various characteristics of electromagnetic and gravitational interactions can be described uniformly, and some equations set of electromagnetic-gravitational field can be achieved.

The base $\mathcal{E}_g$ of the local curved quaternion space (G space, for short) of the gravitational interaction is independent of the base $\mathcal{E}_e$ in E space. Selecting $\mathcal{E}_g = (\mathbf{1}, \vec{i}, \vec{j}, \vec{k}) \circ \vec{e} = (\vec{e}, \vec{I}, \vec{J}, \vec{K})$. So the base $\mathcal{E}_e$ of E space and the base $\mathcal{E}_g$ of G space constitute a base $\mathcal{E}$ of the local curved octonion space.

$$\mathcal{E} = \mathcal{E}_e + \mathcal{E}_g = (\mathbf{1}, \vec{i}, \vec{j}, \vec{k}, \vec{e}, \vec{I}, \vec{J}, \vec{K}) \quad (11)$$

The displacement ($r^0$, $r^1$, $r^2$, $r^3$, $r^4$, $r^5$, $r^6$, $r^7$) in the curved octonion space is

$$\mathcal{R} = (r^0 + \vec{i}\, r^1 + \vec{j}\, r^2 + \vec{k}\, r^3) + (\vec{e}\, r^4 + \vec{I}\, r^5 + \vec{J}\, r^6 + \vec{K}\, r^7) \quad (12)$$

where, $r^0 = ct$, $r^4 = cT$. c is the speed of light beam, t and T denote the time.

Table 3.  The octonion multiplication table

|   | 1 | $\vec{i}$ | $\vec{j}$ | $\vec{k}$ | $\vec{e}$ | $\vec{I}$ | $\vec{J}$ | $\vec{K}$ |
|---|---|---|---|---|---|---|---|---|
| 1 | 1 | $\vec{i}$ | $\vec{j}$ | $\vec{k}$ | $\vec{e}$ | $\vec{I}$ | $\vec{J}$ | $\vec{K}$ |

| $\vec{i}$ | $\vec{i}$ | -1 | $\vec{k}$ | $-\vec{j}$ | $\vec{I}$ | $-\vec{e}$ | $-\vec{K}$ | $\vec{J}$ |
|---|---|---|---|---|---|---|---|---|
| $\vec{j}$ | $\vec{j}$ | $-\vec{k}$ | -1 | $\vec{i}$ | $\vec{J}$ | $\vec{K}$ | $-\vec{e}$ | $-\vec{I}$ |
| $\vec{k}$ | $\vec{k}$ | $\vec{j}$ | $-\vec{i}$ | -1 | $\vec{K}$ | $-\vec{J}$ | $\vec{I}$ | $-\vec{e}$ |
| $\vec{e}$ | $\vec{e}$ | $-\vec{I}$ | $-\vec{J}$ | $-\vec{K}$ | -1 | $\vec{i}$ | $\vec{j}$ | $\vec{k}$ |
| $\vec{I}$ | $\vec{I}$ | $\vec{e}$ | $-\vec{K}$ | $\vec{J}$ | $-\vec{i}$ | -1 | $-\vec{k}$ | $\vec{j}$ |
| $\vec{J}$ | $\vec{J}$ | $\vec{K}$ | $\vec{e}$ | $-\vec{I}$ | $-\vec{j}$ | $\vec{k}$ | -1 | $-\vec{i}$ |
| $\vec{K}$ | $\vec{K}$ | $-\vec{J}$ | $\vec{I}$ | $\vec{e}$ | $-\vec{k}$ | $-\vec{j}$ | $\vec{i}$ | -1 |

The octonion differential operator $\Diamond$ and its conjugate operator $\Diamond^*$ are defined as

$$\Diamond = \Diamond_e + \Diamond_g \quad , \quad \Diamond^* = \Diamond^*_e + \Diamond^*_g \tag{13}$$

where, $\Diamond_g = \vec{e}\nabla_4 + \vec{I}\nabla_5 + \vec{J}\nabla_6 + \vec{K}\nabla_7$ .

The octonion differential operator meets ($\mathcal{Q}$ is an octonion physical quantity)

$$\Diamond^* \circ (\Diamond \circ \mathcal{Q}) = (\Diamond^* \circ \Diamond) \circ \mathcal{Q} = (\Diamond \circ \Diamond^*) \circ \mathcal{Q} \tag{14}$$

In the curved octonion space, the field potential ($a^0, a^1, a^2, a^3, a^4, a^5, a^6, a^7$) of the electromagnetic-gravitational field is defined as

$$\mathcal{A} = \Diamond^* \circ \mathcal{X}$$
$$= (a^0 + \vec{i}a^1 + \vec{j}a^2 + \vec{k}a^3) + (\vec{e}a^4 + \vec{I}a^5 + \vec{J}a^6 + \vec{K}a^7) \tag{15}$$

where, $k_{rx}\mathcal{X} = k^e_{rx}\mathcal{X}_e + k^g_{rx}\mathcal{X}_g$ ; $\mathcal{X}_g$ is the physical quantity in G space.

3.1 Field strength

The field strength $\mathcal{B}$ can be defined as (i, j, k = 0, 1, 2, 3, 4, 5, 6, 7)

$$\begin{aligned}
\mathcal{B} &= \Diamond \circ \mathcal{A} \\
&= b^0 + \vec{i}b^1 + \vec{j}b^2 + \vec{k}b^3 + \vec{e}b^4 + \vec{I}b^5 + \vec{J}b^6 + \vec{K}b^7 \\
&= (\partial_0 a^0 - \partial_1 a^1 - \partial_2 a^2 - \partial_3 a^3 - \partial_4 a^4 - \partial_5 a^5 - \partial_6 a^6 - \partial_7 a^7) \\
&\quad + \vec{i}(\partial_1 a^0 + \partial_0 a^1) + \vec{j}(\partial_2 a^0 + \partial_0 a^2) + \vec{k}(\partial_3 a^0 + \partial_0 a^3) \\
&\quad + \vec{i}(\partial_2 a^3 - \partial_3 a^2) + \vec{j}(\partial_3 a^1 - \partial_1 a^3) + \vec{k}(\partial_1 a^2 - \partial_2 a^1) \\
&\quad + \vec{i}(\partial_4 a^5 - \partial_5 a^4) + \vec{j}(\partial_4 a^6 - \partial_6 a^4) + \vec{k}(\partial_4 a^7 - \partial_7 a^4) \\
&\quad + \vec{i}(\partial_7 a^6 - \partial_6 a^7) + \vec{j}(\partial_5 a^7 - \partial_7 a^5) + \vec{k}(\partial_6 a^5 - \partial_5 a^6) \\
&\quad + \vec{e}\{(\partial_4 a^0 + \partial_0 a^4) + (\partial_5 a^1 - \partial_1 a^5) + (\partial_6 a^2 - \partial_2 a^6) + (\partial_7 a^3 - \partial_3 a^7)\} \\
&\quad + \vec{I}(\partial_5 a^0 + \partial_0 a^5) + \vec{J}(\partial_6 a^0 + \partial_0 a^6) + \vec{K}(\partial_7 a^0 + \partial_0 a^7) \\
&\quad + \vec{I}(\partial_1 a^4 - \partial_4 a^1) + \vec{J}(\partial_2 a^4 - \partial_4 a^2) + \vec{K}(\partial_3 a^4 - \partial_4 a^3) \\
&\quad + \vec{I}(\partial_7 a^2 - \partial_2 a^7) + \vec{J}(\partial_5 a^3 - \partial_3 a^5) + \vec{K}(\partial_6 a^1 - \partial_1 a^6) \\
&\quad + \vec{I}(\partial_3 a^6 - \partial_6 a^3) + \vec{J}(\partial_1 a^7 - \partial_7 a^1) + \vec{K}(\partial_2 a^5 - \partial_5 a^2) \\
&\quad + \Sigma\,(\Gamma_{j0}^{\ 0} - \Gamma_{j1}^{\ 1} - \Gamma_{j2}^{\ 2} - \Gamma_{j3}^{\ 3} - \Gamma_{j4}^{\ 4} - \Gamma_{j5}^{\ 5} - \Gamma_{j6}^{\ 6} - \Gamma_{j7}^{\ 7})\,a^j \\
&\quad + \Sigma\,\{\vec{i}(\Gamma_{j1}^{\ 0} + \Gamma_{j0}^{\ 1}) + \vec{j}(\Gamma_{j2}^{\ 0} + \Gamma_{j0}^{\ 2}) + \vec{k}(\Gamma_{j3}^{\ 0} + \Gamma_{j0}^{\ 3})\}\,a^j \\
&\quad + \Sigma\,\{\vec{i}(\Gamma_{j2}^{\ 3} - \Gamma_{j3}^{\ 2}) + \vec{j}(\Gamma_{j3}^{\ 1} - \Gamma_{j1}^{\ 3}) + \vec{k}(\Gamma_{j1}^{\ 2} - \Gamma_{j2}^{\ 1})\}\,a^j \\
&\quad + \Sigma\,\{\vec{i}(\Gamma_{j4}^{\ 5} - \Gamma_{j5}^{\ 4}) + \vec{j}(\Gamma_{j4}^{\ 6} - \Gamma_{j6}^{\ 4}) + \vec{k}(\Gamma_{j4}^{\ 7} - \Gamma_{j7}^{\ 4})\}\,a^j \\
&\quad + \Sigma\,\{\vec{i}(\Gamma_{j7}^{\ 6} - \Gamma_{j6}^{\ 7}) + \vec{j}(\Gamma_{j5}^{\ 7} - \Gamma_{j7}^{\ 5}) + \vec{k}(\Gamma_{j6}^{\ 5} - \Gamma_{j5}^{\ 6})\}\,a^j \\
&\quad + \vec{e}\,\Sigma\{(\Gamma_{j4}^{\ 0} + \Gamma_{j0}^{\ 4}) + (\Gamma_{j5}^{\ 1} - \Gamma_{j1}^{\ 5}) + (\Gamma_{j6}^{\ 2} - \Gamma_{j2}^{\ 6}) + (\Gamma_{j7}^{\ 3} - \Gamma_{j3}^{\ 7})\}a^j \\
&\quad + \Sigma\,\{\vec{I}(\Gamma_{j5}^{\ 0} + \Gamma_{j0}^{\ 5}) + \vec{J}(\Gamma_{j6}^{\ 0} + \Gamma_{j0}^{\ 6}) + \vec{K}(\Gamma_{j7}^{\ 0} + \Gamma_{j0}^{\ 7})\}\,a^j \\
&\quad + \Sigma\,\{\vec{I}(\Gamma_{j1}^{\ 4} - \Gamma_{j4}^{\ 1}) + \vec{J}(\Gamma_{j2}^{\ 4} - \Gamma_{j4}^{\ 2}) + \vec{K}(\Gamma_{j3}^{\ 4} - \Gamma_{j4}^{\ 3})\}\,a^j \\
&\quad + \Sigma\,\{\vec{I}(\Gamma_{j7}^{\ 2} - \Gamma_{j2}^{\ 7}) + \vec{J}(\Gamma_{j5}^{\ 3} - \Gamma_{j3}^{\ 5}) + \vec{K}(\Gamma_{j6}^{\ 1} - \Gamma_{j1}^{\ 6})\}\,a^j
\end{aligned}$$

$$+ \Sigma \{ \vec{I} (\Gamma_{j3}{}^6 - \Gamma_{j6}{}^3) + \vec{J}(\Gamma_{j1}{}^7 - \Gamma_{j7}{}^1) + \vec{K} (\Gamma_{j2}{}^5 - \Gamma_{j5}{}^2)\} a^j \tag{16}$$

In the above equation, selecting the gauge equation $b^0 = b^4 = 0$ can simplify the definition of field strength in the curved octonion space.

## 3.2 Field source

The field source $S$ ($s^0$, $s^1$, $s^2$, $s^3$, $s^4$, $s^5$, $s^6$, $s^7$) in the curved octonion space is defined as

$$\mu S = (\Diamond + \mathcal{B}/c)^* \circ \mathcal{B}$$
$$= (\Diamond^* \circ \Diamond) \circ \mathcal{A} + \mathcal{B}^* \circ \mathcal{B}/c \tag{17}$$

where, $\Diamond^2 = \Diamond^* \circ \Diamond$. $\mu$ is a coefficient.

The operator $\Diamond^2$ in the curved octonion space can be written as

$$\Diamond^2 = \Diamond^* \circ \Diamond$$
$$= (\nabla_0 - \vec{i}\nabla_1 - \vec{j}\nabla_2 - \vec{k}\nabla_3 - \vec{e}\nabla_4 - \vec{I}\nabla_5 - \vec{J}\nabla_6 - \vec{K}\nabla_7)$$
$$\circ (\nabla_0 + \vec{i}\nabla_1 + \vec{j}\nabla_2 + \vec{k}\nabla_3 + \vec{e}\nabla_4 + \vec{I}\nabla_5 + \vec{J}\nabla_6 + \vec{K}\nabla_7)$$
$$= \nabla_0\nabla_0 + \nabla_1\nabla_1 + \nabla_2\nabla_2 + \nabla_3\nabla_3 + \nabla_4\nabla_4 + \nabla_5\nabla_5 + \nabla_6\nabla_6 + \nabla_7\nabla_7$$
$$+ \vec{i}(\nabla_0\nabla_1 - \nabla_1\nabla_0) + \vec{j}(\nabla_0\nabla_2 - \nabla_2\nabla_0) + \vec{k}(\nabla_0\nabla_3 - \nabla_3\nabla_0)$$
$$+ \vec{i}(\nabla_3\nabla_2 - \nabla_2\nabla_3) + \vec{j}(\nabla_1\nabla_3 - \nabla_3\nabla_1) + \vec{k}(\nabla_2\nabla_1 - \nabla_1\nabla_2)$$
$$+ \vec{i}(\nabla_5\nabla_4 - \nabla_4\nabla_5) + \vec{j}(\nabla_6\nabla_4 - \nabla_4\nabla_6) + \vec{k}(\nabla_7\nabla_4 - \nabla_4\nabla_7)$$
$$+ \vec{i}(\nabla_6\nabla_7 - \nabla_7\nabla_6) + \vec{j}(\nabla_7\nabla_5 - \nabla_5\nabla_7) + \vec{k}(\nabla_5\nabla_6 - \nabla_6\nabla_5)$$
$$+ \vec{e}(\nabla_0\nabla_4 - \nabla_4\nabla_0) + \vec{e}(\nabla_1\nabla_5 - \nabla_5\nabla_1)$$
$$+ \vec{e}(\nabla_2\nabla_6 - \nabla_6\nabla_2) + \vec{e}(\nabla_3\nabla_7 - \nabla_7\nabla_3)$$
$$+ \vec{I}(\nabla_0\nabla_5 - \nabla_5\nabla_0) + \vec{J}(\nabla_0\nabla_6 - \nabla_6\nabla_0) + \vec{K}(\nabla_0\nabla_7 - \nabla_7\nabla_0)$$
$$+ \vec{I}(\nabla_4\nabla_1 - \nabla_1\nabla_4) + \vec{J}(\nabla_7\nabla_1 - \nabla_1\nabla_7) + \vec{K}(\nabla_1\nabla_6 - \nabla_6\nabla_1)$$
$$+ \vec{I}(\nabla_2\nabla_7 - \nabla_7\nabla_2) + \vec{J}(\nabla_4\nabla_2 - \nabla_2\nabla_4) + \vec{K}(\nabla_5\nabla_2 - \nabla_2\nabla_5)$$
$$+ \vec{I}(\nabla_6\nabla_3 - \nabla_3\nabla_6) + \vec{J}(\nabla_3\nabla_5 - \nabla_5\nabla_3) + \vec{K}(\nabla_4\nabla_3 - \nabla_3\nabla_4) \tag{18}$$

In the above equation,

$$(\nabla_m \nabla_k - \nabla_k \nabla_m) a^i = \Sigma R_{kmp}{}^i a^p \tag{19}$$
$$R_{kmp}{}^i = \partial_m \Gamma_{pk}{}^i - \partial_k \Gamma_{pm}{}^i + \Sigma \Gamma_{qm}{}^i \Gamma_{pk}{}^q - \Sigma \Gamma_{qk}{}^i \Gamma_{pm}{}^q \tag{20}$$

wherein, $R_{kmp}{}^i$ is the curvature of curved octonion space; $m$, $n$, $p$, $q = 0, 1, 2, 3, 4, 5, 6, 7$.

Then the definition of field source in the curved octonion space can be rewritten as

$$\mu S = (\Diamond + \mathcal{B}/c)^* \circ \mathcal{B} = (\Diamond^* \circ \Diamond) \circ \mathcal{A} + \mathcal{B}^* \circ \mathcal{B}/c$$
$$= \{ \nabla_0\nabla_0 + \nabla_1\nabla_1 + \nabla_2\nabla_2 + \nabla_3\nabla_3 + \nabla_4\nabla_4 + \nabla_5\nabla_5 + \nabla_6\nabla_6 + \nabla_7\nabla_7$$
$$+ \vec{i}(\nabla_0\nabla_1 - \nabla_1\nabla_0) + \vec{j}(\nabla_0\nabla_2 - \nabla_2\nabla_0) + \vec{k}(\nabla_0\nabla_3 - \nabla_3\nabla_0)$$
$$+ \vec{i}(\nabla_3\nabla_2 - \nabla_2\nabla_3) + \vec{j}(\nabla_1\nabla_3 - \nabla_3\nabla_1) + \vec{k}(\nabla_2\nabla_1 - \nabla_1\nabla_2)$$
$$+ \vec{i}(\nabla_5\nabla_4 - \nabla_4\nabla_5) + \vec{j}(\nabla_6\nabla_4 - \nabla_4\nabla_6) + \vec{k}(\nabla_7\nabla_4 - \nabla_4\nabla_7)$$
$$+ \vec{i}(\nabla_6\nabla_7 - \nabla_7\nabla_6) + \vec{j}(\nabla_7\nabla_5 - \nabla_5\nabla_7) + \vec{k}(\nabla_5\nabla_6 - \nabla_6\nabla_5)$$
$$+ \vec{e}(\nabla_0\nabla_4 - \nabla_4\nabla_0) + \vec{e}(\nabla_1\nabla_5 - \nabla_5\nabla_1)$$
$$+ \vec{e}(\nabla_2\nabla_6 - \nabla_6\nabla_2) + \vec{e}(\nabla_3\nabla_7 - \nabla_7\nabla_3)$$
$$+ \vec{I}(\nabla_0\nabla_5 - \nabla_5\nabla_0) + \vec{J}(\nabla_0\nabla_6 - \nabla_6\nabla_0) + \vec{K}(\nabla_0\nabla_7 - \nabla_7\nabla_0)$$
$$+ \vec{I}(\nabla_4\nabla_1 - \nabla_1\nabla_4) + \vec{J}(\nabla_7\nabla_1 - \nabla_1\nabla_7) + \vec{K}(\nabla_1\nabla_6 - \nabla_6\nabla_1)$$
$$+ \vec{I}(\nabla_2\nabla_7 - \nabla_7\nabla_2) + \vec{J}(\nabla_4\nabla_2 - \nabla_2\nabla_4) + \vec{K}(\nabla_5\nabla_2 - \nabla_2\nabla_5)$$
$$+ \vec{I}(\nabla_6\nabla_3 - \nabla_3\nabla_6) + \vec{J}(\nabla_3\nabla_5 - \nabla_5\nabla_3) + \vec{K}(\nabla_4\nabla_3 - \nabla_3\nabla_4) \} \circ$$
$$(a^0 + \vec{i}a^1 + \vec{j}a^2 + \vec{k}a^3 + \vec{e}a^4 + \vec{I}a^5 + \vec{J}a^6 + \vec{K}a^7) + \mathcal{B}^* \circ \mathcal{B}/c \tag{21}$$

From the above equation, we can obtain the field equations $R_{kmp}{}^i = f(\mu S, \mathcal{B}, \mathcal{A})$, which are much more complex than the field equations in the curved quaternion space or the curved Four-spacetime of Einsteinian General Relativity. [3]

In the curved octonion space, other physical quantities in the electromagnetic- gravitational field including coefficient $\Gamma_{pk}{}^i$ and/or curvature $R_{kmp}{}^i$ can be obtained in the Table 4.

Table 4. The curvature of curved octonion space with other physical quantities

|   | physical quantity | coefficient, $\Gamma_{pk}{}^i$ | curvature, $R_{kmp}{}^i$ |
|---|---|---|---|
| 1 | $\mathcal{X}$ | $\mathcal{A} = \lozenge^* \circ \mathcal{X}$ | $\mathcal{B} = \lozenge^2 \circ \mathcal{X}$ |
| 2 | $\mathcal{A}$ | $\mathcal{B} = \lozenge \circ \mathcal{A}$ | $\mu S = \lozenge^2 \circ \mathcal{A} + \mathcal{B}^* \circ \mathcal{B}/c$ |
| 3 | $\mathcal{B}$ | $\mu S = (\lozenge + \mathcal{B}/c)^* \circ \mathcal{B}$ | $\mu Z = c(\lozenge + \mathcal{B}/c)^2 \circ \mathcal{B}$ |
| 4 | $\mathcal{M}$ | $\mathcal{W} = (c\lozenge + \mathcal{B})^* \circ \mathcal{M}$ | $\mathcal{N} = (c\lozenge + \mathcal{B})^2 \circ \mathcal{M}$ |
| 5 | $\mathcal{B}$ | $\mathcal{T} = (\lozenge + \mathcal{W}/c\hbar)^* \circ \mathcal{B}$ | $\mathcal{O} = (\lozenge + \mathcal{W}/c\hbar)^2 \circ \mathcal{B}$ |
| 6 | $\mathcal{M}$ | $\mathcal{U} = (\lozenge + \mathcal{W}/c\hbar)^* \circ \mathcal{M}$ | $\mathcal{L} = (\lozenge + \mathcal{W}/c\hbar)^2 \circ \mathcal{M}$ |

3.3 Other fields in curved octonion space

With the similar method, we can obtain respectively the field equations of hyper-strong field, strong-weak field and hyper-weak field in the curved octonion space. [4, 5]

In the curved octonion space, other physical quantities in the hyper-strong field including coefficient $\Gamma_{pk}{}^i$ and/or curvature $R_{kmp}{}^i$ can be obtained in the Table 5.

Table 5. The curvature and other physical quantities of the hyper-strong field

|   | physical quantity | coefficient, $\Gamma_{pk}{}^i$ | curvature, $R_{kmp}{}^i$ |
|---|---|---|---|
| 1 | $\mathcal{X}$ | $\mathcal{A} = (\lozenge + \mathcal{X}/k)^* \circ \mathcal{X}$ | $\mathcal{B} = (\lozenge + \mathcal{X}/k)^2 \circ \mathcal{X}$ |
| 2 | $\mathcal{A}$ | $\mathcal{B} = (\lozenge + \mathcal{X}/k) \circ \mathcal{A}$ | $\mu S = (\lozenge + \mathcal{X}/k)^2 \circ \mathcal{A}$ |
| 3 | $\mathcal{B}$ | $\mu S = (\lozenge + \mathcal{X}/k)^* \circ \mathcal{B}$ | $\mu Z = k(\lozenge + \mathcal{X}/k)^2 \circ \mathcal{B}$ |
| 4 | $\mathcal{M}$ | $\mathcal{W} = (k\lozenge + \mathcal{X})^* \circ \mathcal{M}$ | $\mathcal{N} = (k\lozenge + \mathcal{X})^2 \circ \mathcal{M}$ |
| 5 | $\mathcal{X}$ | $\mathcal{D} = (\lozenge + \mathcal{W}/kb)^* \circ \mathcal{X}$ | $\mathcal{G} = (\lozenge + \mathcal{W}/kb)^2 \circ \mathcal{X}$ |
| 6 | $\mathcal{D}$ | $\mathcal{G} = (\lozenge + \mathcal{W}/kb) \circ \mathcal{D}$ | $\mathcal{T} = (\lozenge + \mathcal{W}/kb)^2 \circ \mathcal{D}$ |
| 7 | $\mathcal{G}$ | $\mathcal{T} = (\lozenge + \mathcal{W}/kb)^* \circ \mathcal{G}$ | $\mathcal{O} = (\lozenge + \mathcal{W}/kb)^2 \circ \mathcal{G}$ |
| 8 | $\mathcal{M}$ | $\mathcal{U} = (\lozenge + \mathcal{W}/kb)^* \circ \mathcal{M}$ | $\mathcal{L} = (\lozenge + \mathcal{W}/kb)^2 \circ \mathcal{M}$ |

In the curved octonion space, other physical quantities in the strong-weak field including coefficient $\Gamma_{pk}{}^i$ and/or curvature $R_{kmp}{}^i$ can be obtained in the Table 6.

Table 6. The curvature and other physical quantities of the strong-weak field

|   | physical quantity | coefficient, $\Gamma_{pk}{}^i$ | curvature, $R_{kmp}{}^i$ |
|---|---|---|---|
| 1 | $\mathcal{X}$ | $\mathcal{A} = \lozenge^* \circ \mathcal{X}$ | $\mathcal{B} = \lozenge^2 \circ \mathcal{X}$ |

| 2 | $\mathcal{A}$ | $\mathcal{B} = (\lozenge + \mathcal{A}/k) \circ \mathcal{A}$ | $\mu \mathcal{S} = (\lozenge + \mathcal{A}/k)^2 \circ \mathcal{A}$ |
|---|---|---|---|
| 3 | $\mathcal{B}$ | $\mu \mathcal{S} = (\lozenge + \mathcal{A}/k)^* \circ \mathcal{B}$ | $\mu \mathcal{Z} = k\,(\lozenge + \mathcal{A}/k)^2 \circ \mathcal{B}$ |
| 4 | $\mathcal{M}$ | $\mathcal{W} = (k\lozenge + \mathcal{A})^* \circ \mathcal{M}$ | $\mathcal{N} = (k\lozenge + \mathcal{A})^2 \circ \mathcal{M}$ |
| 5 | $\mathcal{A}$ | $\mathcal{G} = (\lozenge + \mathcal{W}/cb) \circ \mathcal{A}$ | $\mathcal{T} = (\lozenge + \mathcal{W}/cb)^2 \circ \mathcal{A}$ |
| 6 | $\mathcal{G}$ | $\mathcal{T} = (\lozenge + \mathcal{W}/cb)^* \circ \mathcal{G}$ | $\mathcal{O} = (\lozenge + \mathcal{W}/cb)^2 \circ \mathcal{G}$ |
| 7 | $\mathcal{M}$ | $\mathcal{U} = (\lozenge + \mathcal{W}/cb)^* \circ \mathcal{M}$ | $\mathcal{L} = (\lozenge + \mathcal{W}/cb)^2 \circ \mathcal{M}$ |

In the curved octonion space, other physical quantities in the hyper-weak field including coefficient $\Gamma_{pk}{}^i$ and/or curvature $R_{kmp}{}^i$ can be obtained in the Table 7.

Table 7. The curvature and other physical quantities of the hyper-weak field

|   | physical quantity | coefficient, $\Gamma_{pk}{}^i$ | curvature, $R_{kmp}{}^i$ |
|---|---|---|---|
| 1 | $\mathcal{X}$ | $\mathcal{A} = \lozenge^* \circ \mathcal{X}$ | $\mathcal{B} = \lozenge^2 \circ \mathcal{X}$ |
| 2 | $\mathcal{A}$ | $\mathcal{B} = \lozenge \circ \mathcal{A}$ | $\mu \mathcal{S} = \lozenge^2 \circ \mathcal{A}$ |
| 3 | $\mathcal{B}$ | $\mu \mathcal{S} = \lozenge^* \circ \mathcal{B}$ | $\mu \mathcal{Z} = k\,\lozenge^2 \circ \mathcal{B} + \mu \mathcal{S} \circ \mathcal{S}$ |
| 4 | $\mathcal{M}$ | $\mathcal{W} = (k\lozenge + \mathcal{S}/k)^* \circ \mathcal{M}$ | $\mathcal{N} = (k\lozenge + \mathcal{S}/k)^2 \circ \mathcal{M}$ |
| 5 | $\mathcal{M}$ | $\mathcal{U} = (\lozenge + \mathcal{W}/kb)^* \circ \mathcal{M}$ | $\mathcal{L} = (\lozenge + \mathcal{W}/kb)^2 \circ \mathcal{M}$ |

## 4. EG-SW Field equations in curved sedenion space

The electromagnetic-gravitational field and strong-weak field can be described with straight sedenion space [6]. In the paper, the spacetime should be expanded from the straight sedenion space to the curved sedenion space. In the curved sedenion space, some properties of the strong, weak, electromagnetic and gravitational interactions can be described uniformly.

Table 8.  Sedenion multiplication table

| × | 1 | $e_1$ | $e_2$ | $e_3$ | $e_4$ | $e_5$ | $e_6$ | $e_7$ | $e_8$ | $e_9$ | $e_{10}$ | $e_{11}$ | $e_{12}$ | $e_{13}$ | $e_{14}$ | $e_{15}$ |
|---|---|---|---|---|---|---|---|---|---|---|---|---|---|---|---|---|
| **1** | 1 | $e_1$ | $e_2$ | $e_3$ | $e_4$ | $e_5$ | $e_6$ | $e_7$ | $e_8$ | $e_9$ | $e_{10}$ | $e_{11}$ | $e_{12}$ | $e_{13}$ | $e_{14}$ | $e_{15}$ |
| $e_1$ | $e_1$ | -1 | $e_3$ | $-e_2$ | $e_5$ | $-e_4$ | $-e_7$ | $e_6$ | $e_9$ | $-e_8$ | $-e_{11}$ | $e_{10}$ | $-e_{13}$ | $e_{12}$ | $e_{15}$ | $-e_{14}$ |
| $e_2$ | $e_2$ | $-e_3$ | -1 | $e_1$ | $e_6$ | $e_7$ | $-e_4$ | $-e_5$ | $e_{10}$ | $e_{11}$ | $-e_8$ | $-e_9$ | $-e_{14}$ | $-e_{15}$ | $e_{12}$ | $e_{13}$ |
| $e_3$ | $e_3$ | $e_2$ | $-e_1$ | -1 | $e_7$ | $-e_6$ | $e_5$ | $-e_4$ | $e_{11}$ | $-e_{10}$ | $e_9$ | $-e_8$ | $-e_{15}$ | $e_{14}$ | $-e_{13}$ | $e_{12}$ |
| $e_4$ | $e_4$ | $-e_5$ | $-e_6$ | $-e_7$ | -1 | $e_1$ | $e_2$ | $e_3$ | $e_{12}$ | $e_{13}$ | $e_{14}$ | $e_{15}$ | $-e_8$ | $-e_9$ | $-e_{10}$ | $-e_{11}$ |
| $e_5$ | $e_5$ | $e_4$ | $-e_7$ | $e_6$ | $-e_1$ | -1 | $-e_3$ | $e_2$ | $e_{13}$ | $-e_{12}$ | $e_{15}$ | $-e_{14}$ | $e_9$ | $-e_8$ | $e_{11}$ | $-e_{10}$ |
| $e_6$ | $e_6$ | $e_7$ | $e_4$ | $-e_5$ | $-e_2$ | $e_3$ | -1 | $-e_1$ | $e_{14}$ | $-e_{15}$ | $-e_{12}$ | $e_{13}$ | $e_{10}$ | $-e_{11}$ | $-e_8$ | $e_9$ |
| $e_7$ | $e_7$ | $-e_6$ | $e_5$ | $e_4$ | $-e_3$ | $-e_2$ | $e_1$ | -1 | $e_{15}$ | $e_{14}$ | $-e_{13}$ | $-e_{12}$ | $e_{11}$ | $e_{10}$ | $-e_9$ | $-e_8$ |
| $e_8$ | $e_8$ | $-e_9$ | $-e_{10}$ | $-e_{11}$ | $-e_{12}$ | $-e_{13}$ | $-e_{14}$ | $-e_{15}$ | -1 | $e_1$ | $e_2$ | $e_3$ | $e_4$ | $e_5$ | $e_6$ | $e_7$ |
| $e_9$ | $e_9$ | $e_8$ | $-e_{11}$ | $e_{10}$ | $-e_{13}$ | $e_{12}$ | $e_{15}$ | $-e_{14}$ | $-e_1$ | -1 | $-e_3$ | $e_2$ | $-e_5$ | $e_4$ | $e_7$ | $-e_6$ |
| $e_{10}$ | $e_{10}$ | $e_{11}$ | $e_8$ | $-e_9$ | $-e_{14}$ | $-e_{15}$ | $e_{12}$ | $e_{13}$ | $-e_2$ | $e_3$ | -1 | $-e_1$ | $-e_6$ | $-e_7$ | $e_4$ | $e_5$ |
| $e_{11}$ | $e_{11}$ | $-e_{10}$ | $e_9$ | $e_8$ | $-e_{15}$ | $e_{14}$ | $-e_{13}$ | $e_{12}$ | $-e_3$ | $-e_2$ | $e_1$ | -1 | $-e_7$ | $e_6$ | $-e_5$ | $e_4$ |

| $e_{12}$ | $e_{12}$ | $e_{13}$ | $e_{14}$ | $e_{15}$ | $e_8$ | $-e_9$ | $-e_{10}$ | $-e_{11}$ | $-e_4$ | $e_5$ | $e_6$ | $e_7$ | $-1$ | $-e_1$ | $-e_2$ | $-e_3$ |
|---|---|---|---|---|---|---|---|---|---|---|---|---|---|---|---|---|
| $e_{13}$ | $e_{13}$ | $-e_{12}$ | $e_{15}$ | $-e_{14}$ | $e_9$ | $e_8$ | $e_{11}$ | $-e_{10}$ | $-e_5$ | $-e_4$ | $e_7$ | $-e_6$ | $e_1$ | $-1$ | $e_3$ | $-e_2$ |
| $e_{14}$ | $e_{14}$ | $-e_{15}$ | $-e_{12}$ | $e_{13}$ | $e_{10}$ | $-e_{11}$ | $e_8$ | $e_9$ | $-e_6$ | $-e_7$ | $-e_4$ | $e_5$ | $e_2$ | $-e_3$ | $-1$ | $e_1$ |
| $e_{15}$ | $e_{15}$ | $e_{14}$ | $-e_{13}$ | $-e_{12}$ | $e_{11}$ | $e_{10}$ | $-e_9$ | $e_8$ | $-e_7$ | $e_6$ | $-e_5$ | $-e_4$ | $e_3$ | $e_2$ | $-e_1$ | $-1$ |

In the local curved sedenion space EG-SW, the base $\mathcal{E}_{EG\text{-}SW}$ can be written as

$$\mathcal{E}_{EG\text{-}SW} = (1, \vec{e}_1, \vec{e}_2, \vec{e}_3, \vec{e}_4, \vec{e}_5, \vec{e}_6, \vec{e}_7,$$
$$\vec{e}_8, \vec{e}_9, \vec{e}_{10}, \vec{e}_{11}, \vec{e}_{12}, \vec{e}_{13}, \vec{e}_{14}, \vec{e}_{15}) \tag{22}$$

The displacement ( $r_0$, $r_1$, $r_2$, $r_3$, $r_4$, $r_5$, $r_6$, $r_7$, $r_8$, $r_9$, $r_{10}$, $r_{11}$, $r_{12}$, $r_{13}$, $r_{14}$, $r_{15}$ ) in the curved sedenion space EG-SW is

$$\mathcal{R}_{EG\text{-}SW} = r_0 + \vec{e}_1 r_1 + \vec{e}_2 r_2 + \vec{e}_3 r_3 + \vec{e}_4 r_4 + \vec{e}_5 r_5$$
$$+ \vec{e}_6 r_6 + \vec{e}_7 r_7 + \vec{e}_8 r_8 + \vec{e}_9 r_9 + \vec{e}_{10} r_{10}$$
$$+ \vec{e}_{11} r_{11} + \vec{e}_{12} r_{12} + \vec{e}_{13} r_{13} + \vec{e}_{14} r_{14} + \vec{e}_{15} r_{15} \tag{23}$$

In curved sedenion space, the differential operator should be expanded from $\partial_i$ to $\nabla_i$.

$$\nabla_i A^k = \partial_i A^k + \Sigma \Gamma_{ji}{}^k A^j \tag{24}$$

wherein, $\Gamma_{ji}{}^k$ is a coefficient; $A^k$ is a physical quantity; $\partial_i = \partial/\partial r^i$ ; i , j , k = 0, 1, 2, … , 15.

The sedenion differential operator $\Diamond_{EG\text{-}SW}$ and its conjugate operator are defined as,

$$\Diamond_{EG\text{-}SW} = \nabla_0 + \vec{e}_1 \nabla_1 + \vec{e}_2 \nabla_2 + \vec{e}_3 \nabla_3 + \vec{e}_4 \nabla_4 + \vec{e}_5 \nabla_5$$
$$+ \vec{e}_6 \nabla_6 + \vec{e}_7 \nabla_7 + \vec{e}_8 \nabla_8 + \vec{e}_9 \nabla_9 + \vec{e}_{10} \nabla_{10}$$
$$+ \vec{e}_{11} \nabla_{11} + \vec{e}_{12} \nabla_{12} + \vec{e}_{13} \nabla_{13} + \vec{e}_{14} \nabla_{14} + \vec{e}_{15} \nabla_{15} \tag{25}$$

$$\Diamond^*_{EG\text{-}SW} = \nabla_0 - \vec{e}_1 \nabla_1 - \vec{e}_2 \nabla_2 - \vec{e}_3 \nabla_3 - \vec{e}_4 \nabla_4 - \vec{e}_5 \nabla_5$$
$$- \vec{e}_6 \nabla_6 - \vec{e}_7 \nabla_7 - \vec{e}_8 \nabla_8 - \vec{e}_9 \nabla_9 - \vec{e}_{10} \nabla_{10}$$
$$- \vec{e}_{11} \nabla_{11} - \vec{e}_{12} \nabla_{12} - \vec{e}_{13} \nabla_{13} - \vec{e}_{14} \nabla_{14} - \vec{e}_{15} \nabla_{15} \tag{26}$$

4.1 Field potential and strength

In the curved sedenion field EG-SW, the field potential ($a_0$, $a_1$, $a_2$, $a_3$, $a_4$, $a_5$, $a_6$, $a_7$, $a_8$, $a_9$, $a_{10}$, $a_{11}$, $a_{12}$, $a_{13}$, $a_{14}$, $a_{15}$ ) is defined as

$$\mathcal{A} = \Diamond^*_{EG\text{-}SW} \circ \mathcal{X}$$
$$= a_0 + a_1 \vec{e}_1 + a_2 \vec{e}_2 + a_3 \vec{e}_3 + a_4 \vec{e}_4 + a_5 \vec{e}_5$$
$$+ a_6 \vec{e}_6 + a_7 \vec{e}_7 + a_8 \vec{e}_8 + a_9 \vec{e}_9 + a_{10} \vec{e}_{10}$$
$$+ a_{11} \vec{e}_{11} + a_{12} \vec{e}_{12} + a_{13} \vec{e}_{13} + a_{14} \vec{e}_{14} + a_{15} \vec{e}_{15} \tag{27}$$

where, $k_{rx} \mathcal{X} = k^s_{rx} \mathcal{X}_s + k^w_{rx} \mathcal{X}_w + k^e_{rx} \mathcal{X}_e + k^g_{rx} \mathcal{X}_g$.

The field strength $\mathcal{B}$ of the curved sedenion field EG-SW can be defined as

$$\mathcal{B} = \Diamond_{EG\text{-}SW} \circ \mathcal{A}$$
$$= ( \nabla_0 + \vec{e}_1 \nabla_1 + \vec{e}_2 \nabla_2 + \vec{e}_3 \nabla_3 + \vec{e}_4 \nabla_4 + \vec{e}_5 \nabla_5$$
$$+ \vec{e}_6 \nabla_6 + \vec{e}_7 \nabla_7 + \vec{e}_8 \nabla_8 + \vec{e}_9 \nabla_9 + \vec{e}_{10} \nabla_{10}$$
$$+ \vec{e}_{11} \nabla_{11} + \vec{e}_{12} \nabla_{12} + \vec{e}_{13} \nabla_{13} + \vec{e}_{14} \nabla_{14} + \vec{e}_{15} \nabla_{15} )$$
$$\circ ( a_0 + a_1 \vec{e}_1 + a_2 \vec{e}_2 + a_3 \vec{e}_3 + a_4 \vec{e}_4 + a_5 \vec{e}_5$$
$$+ a_6 \vec{e}_6 + a_7 \vec{e}_7 + a_8 \vec{e}_8 + a_9 \vec{e}_9 + a_{10} \vec{e}_{10}$$
$$+ a_{11} \vec{e}_{11} + a_{12} \vec{e}_{12} + a_{13} \vec{e}_{13} + a_{14} \vec{e}_{14} + a_{15} \vec{e}_{15} ) \tag{28}$$

4.2 Field source

The field source of the curved sedenion field can be defined as

$$\mu S = (\mathcal{B}/K + \diamondsuit_{\text{EG-SW}})^* \circ \mathcal{B}$$
$$= (\mathcal{B}_{\text{E-G}}/k^B_{\text{E-G}} + \mathcal{B}_{\text{S-W}}/k^B_{\text{S-W}} + \diamondsuit_{\text{EG-SW}})^* \circ \mathcal{B} \qquad (29)$$

where, $k^B_{\text{S-W}}$ and $k^B_{\text{E-G}}$ are coefficients.

The operator $\diamondsuit^2_{\text{EG-SW}}$ in the curved sedenion space can be written as

$$\begin{aligned}
\diamondsuit^2_{\text{EG-SW}} &= \diamondsuit^*_{\text{EG-SW}} \circ \diamondsuit_{\text{EG-SW}} \\
&= (\nabla_0 - \vec{e}_1 \nabla_1 - \vec{e}_2 \nabla_2 - \vec{e}_3 \nabla_3 - \vec{e}_4 \nabla_4 - \vec{e}_5 \nabla_5 \\
&\quad - \vec{e}_6 \nabla_6 - \vec{e}_7 \nabla_7 - \vec{e}_8 \nabla_8 - \vec{e}_9 \nabla_9 - \vec{e}_{10} \nabla_{10} \\
&\quad - \vec{e}_{11} \nabla_{11} - \vec{e}_{12} \nabla_{12} - \vec{e}_{13} \nabla_{13} - \vec{e}_{14} \nabla_{14} - \vec{e}_{15} \nabla_{15}) \\
&\quad \circ (\nabla_0 + \vec{e}_1 \nabla_1 + \vec{e}_2 \nabla_2 + \vec{e}_3 \nabla_3 + \vec{e}_4 \nabla_4 + \vec{e}_5 \nabla_5 \\
&\quad + \vec{e}_6 \nabla_6 + \vec{e}_7 \nabla_7 + \vec{e}_8 \nabla_8 + \vec{e}_9 \nabla_9 + \vec{e}_{10} \nabla_{10} \\
&\quad + \vec{e}_{11} \nabla_{11} + \vec{e}_{12} \nabla_{12} + \vec{e}_{13} \nabla_{13} + \vec{e}_{14} \nabla_{14} + \vec{e}_{15} \nabla_{15}) \\
&= \nabla_0 \nabla_0 + \nabla_1 \nabla_1 + \nabla_2 \nabla_2 + \nabla_3 \nabla_3 + \nabla_4 \nabla_4 + \nabla_5 \nabla_5 \\
&\quad + \nabla_6 \nabla_6 + \nabla_7 \nabla_7 + \nabla_8 \nabla_8 + \nabla_9 \nabla_9 + \nabla_{10} \nabla_{10} \\
&\quad + \nabla_{11} \nabla_{11} + \nabla_{12} \nabla_{12} + \nabla_{13} \nabla_{13} + \nabla_{14} \nabla_{14} + \nabla_{15} \nabla_{15} \\
&\quad + \vec{e}_1 (\nabla_0 \nabla_1 - \nabla_1 \nabla_0) + \vec{e}_1 (\nabla_3 \nabla_2 - \nabla_2 \nabla_3) + \vec{e}_1 (\nabla_5 \nabla_4 - \nabla_4 \nabla_5) \\
&\quad + \vec{e}_1 (\nabla_6 \nabla_7 - \nabla_7 \nabla_6) + \vec{e}_1 (\nabla_9 \nabla_8 - \nabla_8 \nabla_9) + \vec{e}_1 (\nabla_{10} \nabla_{11} - \nabla_{11} \nabla_{10}) \\
&\quad + \vec{e}_1 (\nabla_{12} \nabla_{13} - \nabla_{13} \nabla_{12}) + \vec{e}_1 (\nabla_{15} \nabla_{14} - \nabla_{14} \nabla_{15}) \\
&\quad + \vec{e}_2 (\nabla_0 \nabla_2 - \nabla_2 \nabla_0) + \vec{e}_2 (\nabla_1 \nabla_3 - \nabla_3 \nabla_1) + \vec{e}_2 (\nabla_6 \nabla_4 - \nabla_4 \nabla_6) \\
&\quad + \vec{e}_2 (\nabla_7 \nabla_5 - \nabla_5 \nabla_7) + \vec{e}_2 (\nabla_{10} \nabla_8 - \nabla_8 \nabla_{10}) + \vec{e}_2 (\nabla_{11} \nabla_9 - \nabla_9 \nabla_{11}) \\
&\quad + \vec{e}_2 (\nabla_{12} \nabla_{14} - \nabla_{14} \nabla_{12}) + \vec{e}_2 (\nabla_{13} \nabla_{15} - \nabla_{15} \nabla_{13}) \\
&\quad + \vec{e}_3 (\nabla_0 \nabla_3 - \nabla_3 \nabla_0) + \vec{e}_3 (\nabla_2 \nabla_1 - \nabla_1 \nabla_2) + \vec{e}_3 (\nabla_7 \nabla_4 - \nabla_4 \nabla_7) \\
&\quad + \vec{e}_3 (\nabla_5 \nabla_6 - \nabla_6 \nabla_5) + \vec{e}_3 (\nabla_{11} \nabla_8 - \nabla_8 \nabla_{11}) + \vec{e}_3 (\nabla_9 \nabla_{10} - \nabla_{10} \nabla_9) \\
&\quad + \vec{e}_3 (\nabla_{12} \nabla_{15} - \nabla_{15} \nabla_{12}) + \vec{e}_3 (\nabla_{14} \nabla_{13} - \nabla_{13} \nabla_{14}) \\
&\quad + \vec{e}_4 (\nabla_0 \nabla_4 - \nabla_4 \nabla_0) + \vec{e}_4 (\nabla_1 \nabla_5 - \nabla_5 \nabla_1) + \vec{e}_4 (\nabla_2 \nabla_6 - \nabla_6 \nabla_2) \\
&\quad + \vec{e}_4 (\nabla_3 \nabla_7 - \nabla_7 \nabla_3) + \vec{e}_4 (\nabla_{12} \nabla_8 - \nabla_8 \nabla_{12}) + \vec{e}_4 (\nabla_{13} \nabla_9 - \nabla_9 \nabla_{13}) \\
&\quad + \vec{e}_4 (\nabla_{14} \nabla_{10} - \nabla_{10} \nabla_{14}) + \vec{e}_4 (\nabla_{15} \nabla_{11} - \nabla_{11} \nabla_{15}) \\
&\quad + \vec{e}_5 (\nabla_0 \nabla_5 - \nabla_5 \nabla_0) + \vec{e}_5 (\nabla_4 \nabla_1 - \nabla_1 \nabla_4) + \vec{e}_5 (\nabla_2 \nabla_7 - \nabla_7 \nabla_2) \\
&\quad + \vec{e}_5 (\nabla_6 \nabla_3 - \nabla_3 \nabla_6) + \vec{e}_5 (\nabla_{13} \nabla_8 - \nabla_8 \nabla_{13}) + \vec{e}_5 (\nabla_9 \nabla_{12} - \nabla_{12} \nabla_9) \\
&\quad + \vec{e}_5 (\nabla_{15} \nabla_{10} - \nabla_{10} \nabla_{15}) + \vec{e}_5 (\nabla_{11} \nabla_{14} - \nabla_{14} \nabla_{11}) \\
&\quad + \vec{e}_6 (\nabla_0 \nabla_6 - \nabla_6 \nabla_0) + \vec{e}_6 (\nabla_7 \nabla_1 - \nabla_1 \nabla_7) + \vec{e}_6 (\nabla_4 \nabla_2 - \nabla_2 \nabla_4) \\
&\quad + \vec{e}_6 (\nabla_3 \nabla_5 - \nabla_5 \nabla_3) + \vec{e}_6 (\nabla_{14} \nabla_8 - \nabla_8 \nabla_{14}) + \vec{e}_6 (\nabla_9 \nabla_{15} - \nabla_{15} \nabla_9) \\
&\quad + \vec{e}_6 (\nabla_{10} \nabla_{12} - \nabla_{12} \nabla_{10}) + \vec{e}_6 (\nabla_{13} \nabla_{11} - \nabla_{11} \nabla_{13}) \\
&\quad + \vec{e}_7 (\nabla_0 \nabla_7 - \nabla_7 \nabla_0) + \vec{e}_7 (\nabla_1 \nabla_6 - \nabla_6 \nabla_1) + \vec{e}_7 (\nabla_5 \nabla_2 - \nabla_2 \nabla_5) \\
&\quad + \vec{e}_7 (\nabla_4 \nabla_3 - \nabla_3 \nabla_4) + \vec{e}_7 (\nabla_{15} \nabla_8 - \nabla_8 \nabla_{15}) + \vec{e}_7 (\nabla_{14} \nabla_9 - \nabla_9 \nabla_{14}) \\
&\quad + \vec{e}_7 (\nabla_{10} \nabla_{13} - \nabla_{13} \nabla_{10}) + \vec{e}_7 (\nabla_{11} \nabla_{12} - \nabla_{12} \nabla_{11}) \\
&\quad + \vec{e}_8 (\nabla_0 \nabla_8 - \nabla_8 \nabla_0) + \vec{e}_8 (\nabla_1 \nabla_9 - \nabla_9 \nabla_1) + \vec{e}_8 (\nabla_2 \nabla_{10} - \nabla_{10} \nabla_2) \\
&\quad + \vec{e}_8 (\nabla_3 \nabla_{11} - \nabla_{11} \nabla_3) + \vec{e}_8 (\nabla_4 \nabla_{12} - \nabla_{12} \nabla_4) + \vec{e}_8 (\nabla_5 \nabla_{13} - \nabla_{13} \nabla_5) \\
&\quad + \vec{e}_8 (\nabla_6 \nabla_{14} - \nabla_{14} \nabla_6) + \vec{e}_8 (\nabla_7 \nabla_{15} - \nabla_{15} \nabla_7) \\
&\quad + \vec{e}_9 (\nabla_0 \nabla_9 - \nabla_9 \nabla_0) + \vec{e}_9 (\nabla_8 \nabla_1 - \nabla_1 \nabla_8) + \vec{e}_9 (\nabla_2 \nabla_{11} - \nabla_{11} \nabla_2) \\
&\quad + \vec{e}_9 (\nabla_4 \nabla_{13} - \nabla_{13} \nabla_4) + \vec{e}_9 (\nabla_{15} \nabla_6 - \nabla_6 \nabla_{15}) + \vec{e}_9 (\nabla_7 \nabla_{14} - \nabla_{14} \nabla_7) \\
&\quad + \vec{e}_9 (\nabla_{10} \nabla_3 - \nabla_3 \nabla_{10}) + \vec{e}_9 (\nabla_{12} \nabla_5 - \nabla_5 \nabla_{12})
\end{aligned}$$

$+\vec{e}_{10}(\nabla_0\nabla_{10} - \nabla_{10}\nabla_0) + \vec{e}_{10}(\nabla_{11}\nabla_1 - \nabla_1\nabla_{11}) + \vec{e}_{10}(\nabla_8\nabla_2 - \nabla_2\nabla_8)$
$+\vec{e}_{10}(\nabla_3\nabla_9 - \nabla_9\nabla_3) + \vec{e}_{10}(\nabla_4\nabla_{14} - \nabla_{14}\nabla_4) + \vec{e}_{10}(\nabla_5\nabla_{15} - \nabla_{15}\nabla_5)$
$+\vec{e}_{10}(\nabla_{12}\nabla_6 - \nabla_6\nabla_{12}) + \vec{e}_{10}(\nabla_{13}\nabla_7 - \nabla_7\nabla_{13})$
$+\vec{e}_{11}(\nabla_0\nabla_{11} - \nabla_{11}\nabla_0) + \vec{e}_{11}(\nabla_1\nabla_{10} - \nabla_{10}\nabla_1) + \vec{e}_{11}(\nabla_9\nabla_2 - \nabla_2\nabla_9)$
$+\vec{e}_{11}(\nabla_8\nabla_3 - \nabla_3\nabla_8) + \vec{e}_{11}(\nabla_4\nabla_{15} - \nabla_5\nabla_{14}) + \vec{e}_{11}(\nabla_6\nabla_{13} - \nabla_{13}\nabla_6)$
$+\vec{e}_{11}(\nabla_{12}\nabla_7 - \nabla_7\nabla_{12}) + \vec{e}_{11}(\nabla_{14}\nabla_5 - \nabla_{15}\nabla_4)$
$+\vec{e}_{12}(\nabla_0\nabla_{12} - \nabla_{12}\nabla_0) + \vec{e}_{12}(\nabla_{13}\nabla_1 - \nabla_1\nabla_{13}) + \vec{e}_{12}(\nabla_{14}\nabla_2 - \nabla_2\nabla_{14})$
$+\vec{e}_{12}(\nabla_{15}\nabla_3 - \nabla_3\nabla_{15}) + \vec{e}_{12}(\nabla_5\nabla_9 - \nabla_9\nabla_5) + \vec{e}_{12}(\nabla_6\nabla_{10} - \nabla_{10}\nabla_6)$
$+\vec{e}_{12}(\nabla_7\nabla_{11} - \nabla_{11}\nabla_7) + \vec{e}_{12}(\nabla_8\nabla_4 - \nabla_4\nabla_8)$
$+\vec{e}_{13}(\nabla_0\nabla_{13} - \nabla_{13}\nabla_0) + \vec{e}_{13}(\nabla_1\nabla_{12} - \nabla_{12}\nabla_1) + \vec{e}_{13}(\nabla_{15}\nabla_2 - \nabla_2\nabla_{15})$
$+\vec{e}_{13}(\nabla_3\nabla_{14} - \nabla_{14}\nabla_3) + \vec{e}_{13}(\nabla_9\nabla_4 - \nabla_4\nabla_9) + \vec{e}_{13}(\nabla_8\nabla_5 - \nabla_5\nabla_8)$
$+\vec{e}_{13}(\nabla_{11}\nabla_6 - \nabla_6\nabla_{11}) + \vec{e}_{13}(\nabla_7\nabla_{10} - \nabla_{10}\nabla_7)$
$+\vec{e}_{14}(\nabla_0\nabla_{14} - \nabla_{14}\nabla_0) + \vec{e}_{14}(\nabla_1\nabla_{15} - \nabla_{15}\nabla_1) + \vec{e}_{14}(\nabla_2\nabla_{12} - \nabla_{12}\nabla_2)$
$+\vec{e}_{14}(\nabla_{13}\nabla_3 - \nabla_3\nabla_{13}) + \vec{e}_{14}(\nabla_{10}\nabla_4 - \nabla_4\nabla_{10}) + \vec{e}_{14}(\nabla_5\nabla_{11} - \nabla_{11}\nabla_5)$
$+\vec{e}_{14}(\nabla_8\nabla_6 - \nabla_6\nabla_8) + \vec{e}_{14}(\nabla_9\nabla_7 - \nabla_7\nabla_9)$
$+\vec{e}_{15}(\nabla_0\nabla_{15} - \nabla_{15}\nabla_0) + \vec{e}_{15}(\nabla_{14}\nabla_1 - \nabla_1\nabla_{14}) + \vec{e}_{15}(\nabla_2\nabla_{13} - \nabla_{13}\nabla_2)$
$+\vec{e}_{15}(\nabla_3\nabla_{12} - \nabla_{12}\nabla_3) + \vec{e}_{15}(\nabla_{11}\nabla_4 - \nabla_4\nabla_{11}) + \vec{e}_{15}(\nabla_{10}\nabla_5 - \nabla_5\nabla_{10})$
$+\vec{e}_{15}(\nabla_6\nabla_9 - \nabla_9\nabla_6) + \vec{e}_{15}(\nabla_8\nabla_7 - \nabla_7\nabla_8)$ (30)

In the above equation,

$$(\nabla_m\nabla_k - \nabla_k\nabla_m) a^i = \Sigma R_{kmp}{}^i a^p \quad (31)$$

$$R_{kmp}{}^i = \partial_m \Gamma_{pk}{}^i - \partial_k \Gamma_{pm}{}^i + \Sigma \Gamma_{qm}{}^i \Gamma_{pk}{}^q - \Sigma \Gamma_{qk}{}^i \Gamma_{pm}{}^q \quad (32)$$

wherein, $R_{kmp}{}^i$ is the curvature of curved sedenion space; $m, n, p, q = 0, 1, 2, 3, \ldots, 15$.

Then the definition of field source in the curved sedenion space can be rewritten as

$$\mu S = (\mathcal{B}/K + \diamondsuit_{EG-SW})^* \circ \mathcal{B}$$
$$= (\diamondsuit^*_{EG-SW} \circ \diamondsuit_{EG-SW}) \circ \mathcal{A} + (\mathcal{B}_{E-G} / k^B_{E-G} + \mathcal{B}_{S-W} / k^B_{S-W})^* \circ \mathcal{B} \quad (33)$$

From the above equation, we can obtain the field equations $R_{kmp}{}^i = f(\mu S, \mathcal{B}, \mathcal{A}, K)$. In the curved octonion space ($\diamondsuit = \diamondsuit_{EG-SW}$), other physical quantities in the curved sedenion field EG-SW including the coefficient $\Gamma_{pk}{}^i$ and/or curvature $R_{kmp}{}^i$ can be obtained in the Table 9. Where, the coefficient $\mathcal{B}/K = \mathcal{B}_{E-G} / k^B_{E-G} + \mathcal{B}_{S-W} / k^B_{S-W}$, and coefficient $\mathcal{W}/K b = \mathcal{W}_{E-G} / k^B_{E-G} b^B_{E-G} + \mathcal{W}_{S-W} / k^B_{S-W} b^B_{S-W}$.

Table 9. The curvature of curved sedenion space with other physical quantities

|   | physical quantity | coefficient, $\Gamma_{pk}{}^i$ | curvature, $R_{kmp}{}^i$ |
|---|---|---|---|
| 1 | $\mathcal{X}$ | $\mathcal{A} = \diamondsuit^* \circ \mathcal{X}$ | $\mathcal{B} = \diamondsuit^2 \circ \mathcal{X}$ |
| 2 | $\mathcal{A}$ | $\mathcal{B} = \diamondsuit \circ \mathcal{A}$ | $\mu S = \diamondsuit^2 \circ \mathcal{A} + (\mathcal{B}/K)^* \circ \mathcal{B}$ |
| 3 | $\mathcal{B}$ | $\mu S = (\diamondsuit + \mathcal{B}/K)^* \circ \mathcal{B}$ | $\mu Z = K(\diamondsuit + \mathcal{B}/K)^2 \circ \mathcal{B}$ |
| 4 | $\mathcal{M}$ | $\mathcal{W} = K(\diamondsuit + \mathcal{B}/K)^* \circ \mathcal{M}$ | $\mathcal{N} = K^2(\diamondsuit + \mathcal{B}/K)^2 \circ \mathcal{M}$ |
| 5 | $\mathcal{B}$ | $\mathcal{T} = (\diamondsuit + \mathcal{W}/K b)^* \circ \mathcal{B}$ | $O = (\diamondsuit + \mathcal{W}/K b)^2 \circ \mathcal{B}$ |
| 6 | $\mathcal{M}$ | $\mathcal{U} = (\diamondsuit + \mathcal{W}/K b)^* \circ \mathcal{M}$ | $\mathcal{L} = (\diamondsuit + \mathcal{W}/K b)^2 \circ \mathcal{M}$ |

4.3 Other fields in the curved sedenion space

With the similar method, we can obtain respectively the field equations of the sedenion field SW-EG, the sedenion field HS-SW and sedenion field SW-EG in the curved sedenion space.

In the curved sedenion space, other physical quantities in the sedenion field SW-EG including coefficient $\Gamma_{pk}{}^i$ and/or curvature $R_{kmp}{}^i$ can be obtained in Table 10. Where, $\mathcal{A}/K = \mathcal{A}_{E-G}/k^A{}_{E-G} + \mathcal{A}_{S-W}/k^A{}_{S-W}$, and $\mathcal{W}/Kb = \mathcal{W}_{E-G}/k^A{}_{E-G}b^A{}_{E-G} + \mathcal{W}_{S-W}/k^A{}_{S-W}b^A{}_{S-W}$.

Table 10.  The curvature and other physical quantities of the sedenion field SW-EG

|   | physical quantity | coefficient, $\Gamma_{pk}{}^i$ | curvature, $R_{kmp}{}^i$ |
|---|---|---|---|
| 1 | $\mathcal{X}$ | $\mathcal{A} = \lozenge^* \circ \mathcal{X}$ | $\mathcal{B} = \lozenge^2 \circ \mathcal{X}$ |
| 2 | $\mathcal{A}$ | $\mathcal{B} = (\lozenge + \mathcal{A}/K) \circ \mathcal{A}$ | $\mu\mathcal{S} = (\lozenge + \mathcal{A}/K)^2 \circ \mathcal{A}$ |
| 3 | $\mathcal{B}$ | $\mu\mathcal{S} = (\lozenge + \mathcal{A}/K)^* \circ \mathcal{B}$ | $\mu\mathcal{Z} = K(\lozenge + \mathcal{A}/K)^2 \circ \mathcal{B}$ |
| 4 | $\mathcal{M}$ | $\mathcal{W} = K(\lozenge + \mathcal{A}/K)^* \circ \mathcal{M}$ | $\mathcal{N} = K(\lozenge + \mathcal{A}/K)^2 \circ \mathcal{M}$ |
| 5 | $\mathcal{A}$ | $\mathcal{G} = (\lozenge + \mathcal{W}/Kb) \circ \mathcal{A}$ | $\mathcal{T} = (\lozenge + \mathcal{W}/Kb)^2 \circ \mathcal{A}$ |
| 6 | $\mathcal{G}$ | $\mathcal{T} = (\lozenge + \mathcal{W}/Kb)^* \circ \mathcal{G}$ | $\mathcal{O} = (\lozenge + \mathcal{W}/Kb)^2 \circ \mathcal{G}$ |
| 7 | $\mathcal{M}$ | $\mathcal{U} = (\lozenge + \mathcal{W}/Kb)^* \circ \mathcal{M}$ | $\mathcal{L} = (\lozenge + \mathcal{W}/Kb)^2 \circ \mathcal{M}$ |

In the curved sedenion space, other physical quantities in the sedenion field HS-SW including coefficient $\Gamma_{pk}{}^i$ and/or curvature $R_{kmp}{}^i$ can be obtained in Table 11. Where, $\mathcal{X}/K = \mathcal{X}_{H-S}/k^X{}_{H-S} + \mathcal{X}_{S-W}/k^X{}_{S-W}$, and $\mathcal{W}/Kb = \mathcal{W}_{H-S}/k^X{}_{H-S}b^X{}_{H-S} + \mathcal{W}_{S-W}/k^X{}_{S-W}b^X{}_{S-W}$.

Table 11.  The curvature and other physical quantities of the sedenion field HS-SW

|   | physical quantity | coefficient, $\Gamma_{pk}{}^i$ | curvature, $R_{kmp}{}^i$ |
|---|---|---|---|
| 1 | $\mathcal{X}$ | $\mathcal{A} = (\lozenge + \mathcal{X}/K)^* \circ \mathcal{X}$ | $\mathcal{B} = (\lozenge + \mathcal{X}/K)^2 \circ \mathcal{X}$ |
| 2 | $\mathcal{A}$ | $\mathcal{B} = (\lozenge + \mathcal{X}/K) \circ \mathcal{A}$ | $\mu\mathcal{S} = (\lozenge + \mathcal{X}/K)^2 \circ \mathcal{A}$ |
| 3 | $\mathcal{B}$ | $\mu\mathcal{S} = (\lozenge + \mathcal{X}/K)^* \circ \mathcal{B}$ | $\mu\mathcal{Z} = K(\lozenge + \mathcal{X}/K)^2 \circ \mathcal{B}$ |
| 4 | $\mathcal{M}$ | $\mathcal{W} = K(\lozenge + \mathcal{X}/K)^* \circ \mathcal{M}$ | $\mathcal{N} = K(\lozenge + \mathcal{X}/K)^2 \circ \mathcal{M}$ |
| 5 | $\mathcal{X}$ | $\mathcal{D} = (\lozenge + \mathcal{W}/Kb)^* \circ \mathcal{X}$ | $\mathcal{G} = (\lozenge + \mathcal{W}/Kb)^2 \circ \mathcal{X}$ |
| 6 | $\mathcal{D}$ | $\mathcal{G} = (\lozenge + \mathcal{W}/Kb) \circ \mathcal{D}$ | $\mathcal{T} = (\lozenge + \mathcal{W}/Kb)^2 \circ \mathcal{D}$ |
| 7 | $\mathcal{G}$ | $\mathcal{T} = (\lozenge + \mathcal{W}/Kb)^* \circ \mathcal{G}$ | $\mathcal{O} = (\lozenge + \mathcal{W}/Kb)^2 \circ \mathcal{G}$ |
| 8 | $\mathcal{M}$ | $\mathcal{U} = (\lozenge + \mathcal{W}/Kb)^* \circ \mathcal{M}$ | $\mathcal{L} = (\lozenge + \mathcal{W}/Kb)^2 \circ \mathcal{M}$ |

In the curved sedenion space, other physical quantities in the sedenion field SW-HS including coefficient $\Gamma_{pk}{}^i$ and/or curvature $R_{kmp}{}^i$ can be obtained in Table 12. Where, $\mathcal{A}/K = \mathcal{A}_{H-S}/k^A{}_{H-S} + \mathcal{A}_{S-W}/k^A{}_{S-W}$, and $\mathcal{W}/Kb = \mathcal{W}_{H-S}/k^A{}_{H-S}b^A{}_{H-S} + \mathcal{W}_{S-W}/k^A{}_{S-W}b^A{}_{S-W}$.

Table 12.  The curvature and other physical quantities of the sedenion field SW-HS

|   | physical quantity | coefficient, $\Gamma_{pk}{}^i$ | curvature, $R_{kmp}{}^i$ |
|---|---|---|---|
| 1 | $\mathcal{X}$ | $\mathcal{A} = \lozenge^* \circ \mathcal{X}$ | $\mathcal{B} = \lozenge^2 \circ \mathcal{X}$ |
| 2 | $\mathcal{A}$ | $\mathcal{B} = (\lozenge + \mathcal{A}/K) \circ \mathcal{A}$ | $\mu\mathcal{S} = (\lozenge + \mathcal{A}/K)^2 \circ \mathcal{A}$ |
| 3 | $\mathcal{B}$ | $\mu\mathcal{S} = (\lozenge + \mathcal{A}/K)^* \circ \mathcal{B}$ | $\mu\mathcal{Z} = K(\lozenge + \mathcal{A}/K)^2 \circ \mathcal{B}$ |

| 4 | $\mathcal{M}$ | $\mathcal{W} = K (\Diamond + \mathcal{A}/K)^* \circ \mathcal{M}$ | $\mathcal{N} = K (\Diamond + \mathcal{A}/K)^2 \circ \mathcal{M}$ |
| 5 | $\mathcal{A}$ | $\mathcal{G} = (\Diamond + \mathcal{W}/K b) \circ \mathcal{A}$ | $\mathcal{T} = (\Diamond + \mathcal{W}/K b)^2 \circ \mathcal{A}$ |
| 6 | $\mathcal{G}$ | $\mathcal{T} = (\Diamond + \mathcal{W}/K b)^* \circ \mathcal{G}$ | $O = (\Diamond + \mathcal{W}/K b)^2 \circ \mathcal{G}$ |
| 7 | $\mathcal{M}$ | $\mathcal{U} = (\Diamond + \mathcal{W}/K b)^* \circ \mathcal{M}$ | $\mathcal{L} = (\Diamond + \mathcal{W}/K b)^2 \circ \mathcal{M}$ |

## 5. T-B Field equations in curved trigintaduonion space

The electromagnetic-gravitational field, strong-weak field, hyper-strong field and hyper-weak field can be described with the straight trigintaduonion space [7]. In the paper, the spacetime should be expanded from the straight trigintaduonion space to the curved trigintaduonion space. In the curved trigintaduonion space, some properties of four sorts of fields including the hyper-strong field, strong-weak field, electromagnetic-gravitational field and hyper-weak field etc can be described uniformly.

We define the product and conjugate on the trigintaduonions, (u, v) and (x, y), in terms of the sedenions, u, v, x and y, as follows:

$$(u, v)(x, y) = (u x - y^* v, y u + v x^*), \qquad (u, v)^* = (u^*, -v)$$

where, the mark (*) denotes the conjugate.

In the local curved trigintaduonion space T-B, the base $\mathcal{E}_{T-B}$ can be written as

$$\mathcal{E}_{T-B} = (1, \vec{e}_1, \vec{e}_2, \vec{e}_3, \vec{e}_4, \vec{e}_5, \vec{e}_6, \vec{e}_7, \vec{e}_8, \vec{e}_9, \vec{e}_{10},$$
$$\vec{e}_{11}, \vec{e}_{12}, \vec{e}_{13}, \vec{e}_{14}, \vec{e}_{15}, \vec{e}_{16}, \vec{e}_{17}, \vec{e}_{18}, \vec{e}_{19}, \vec{e}_{20},$$
$$\vec{e}_{21}, \vec{e}_{22}, \vec{e}_{23}, \vec{e}_{24}, \vec{e}_{25}, \vec{e}_{26}, \vec{e}_{27}, \vec{e}_{28}, \vec{e}_{29}, \vec{e}_{30}, \vec{e}_{31}) \qquad (34)$$

The displacement $\mathcal{R}_{T-B} = (R_0, R_1, R_2, R_3, R_4, R_5, R_6, R_7, R_8, R_9, R_{10}, R_{11}, R_{12}, R_{13}, R_{14}, R_{15}, R_{16}, R_{17}, R_{18}, R_{19}, R_{20}, R_{21}, R_{22}, R_{23}, R_{24}, R_{25}, R_{26}, R_{27}, R_{28}, R_{29}, R_{30}, R_{31})$ in curved trigintaduonion space T-B is

$$\mathcal{R}_{T-B} = R_0 + \vec{e}_1 R_1 + \vec{e}_2 R_2 + \vec{e}_3 R_3 + \vec{e}_4 R_4 + \vec{e}_5 R_5 + \vec{e}_6 R_6$$
$$+ \vec{e}_7 R_7 + \vec{e}_8 R_8 + \vec{e}_9 R_9 + \vec{e}_{10} R_{10} + \vec{e}_{11} R_{11}$$
$$+ \vec{e}_{12} R_{12} + \vec{e}_{13} R_{13} + \vec{e}_{14} R_{14} + \vec{e}_{15} R_{15} + \vec{e}_{16} R_{16}$$
$$+ \vec{e}_{17} R_{17} + \vec{e}_{18} R_{18} + \vec{e}_{19} R_{19} + \vec{e}_{20} R_{20} + \vec{e}_{21} R_{21}$$
$$+ \vec{e}_{22} R_{22} + \vec{e}_{23} R_{23} + \vec{e}_{24} R_{24} + \vec{e}_{25} R_{25} + \vec{e}_{26} R_{26}$$
$$+ \vec{e}_{27} R_{27} + \vec{e}_{28} R_{28} + \vec{e}_{29} R_{29} + \vec{e}_{30} R_{30} + \vec{e}_{31} R_{31} \qquad (35)$$

In the curved trigintaduonion space, the differential operator should be expanded from $\partial_i$ to $\nabla_i$.

$$\nabla_i A^k = \partial_i A^k + \Sigma \Gamma_{ji}{}^k A^j \qquad (36)$$

wherein, $\Gamma_{ji}{}^k$ is a coefficient; $A^k$ is a physical quantity; $\partial_i = \partial/\partial r^i$; i, j, k = 0, 1, 2, … , 31.

The trigintaduonion differential operator $\Diamond_{T-B}$ and its conjugate operator are defined as,

$$\Diamond_{T-B} = \nabla_0 + \vec{e}_1 \nabla_1 + \vec{e}_2 \nabla_2 + \vec{e}_3 \nabla_3 + \vec{e}_4 \nabla_4 + \vec{e}_5 \nabla_5$$
$$+ \vec{e}_6 \nabla_6 + \vec{e}_7 \nabla_7 + \vec{e}_8 \nabla_8 + \vec{e}_9 \nabla_9 + \vec{e}_{10} \nabla_{10} + \vec{e}_{11} \nabla_{11}$$
$$+ \vec{e}_{12} \nabla_{12} + \vec{e}_{13} \nabla_{13} + \vec{e}_{14} \nabla_{14} + \vec{e}_{15} \nabla_{15} + \vec{e}_{16} \nabla_{16}$$
$$+ \vec{e}_{17} \nabla_{17} + \vec{e}_{18} \nabla_{18} + \vec{e}_{19} \nabla_{19} + \vec{e}_{20} \nabla_{20} + \vec{e}_{21} \nabla_{21}$$
$$+ \vec{e}_{22} \nabla_{22} + \vec{e}_{23} \nabla_{23} + \vec{e}_{24} \nabla_{24} + \vec{e}_{25} \nabla_{25} + \vec{e}_{26} \nabla_{26}$$
$$+ \vec{e}_{27} \nabla_{27} + \vec{e}_{28} \nabla_{28} + \vec{e}_{29} \nabla_{29} + \vec{e}_{30} \nabla_{30} + \vec{e}_{31} \nabla_{31} \qquad (37)$$
$$\Diamond^*_{T-B} = \nabla_0 - \vec{e}_1 \nabla_1 - \vec{e}_2 \nabla_2 - \vec{e}_3 \nabla_3 - \vec{e}_4 \nabla_4 - \vec{e}_5 \nabla_5$$

$$- \vec{e}_6 \nabla_6 - \vec{e}_7 \nabla_7 - \vec{e}_8 \nabla_8 - \vec{e}_9 \nabla_9 - \vec{e}_{10} \nabla_{10} - \vec{e}_{11} \nabla_{11}$$
$$- \vec{e}_{12} \nabla_{12} - \vec{e}_{13} \nabla_{13} - \vec{e}_{14} \nabla_{14} - \vec{e}_{15} \nabla_{15} - \vec{e}_{16} \nabla_{16}$$
$$- \vec{e}_{17} \nabla_{17} - \vec{e}_{18} \nabla_{18} - \vec{e}_{19} \nabla_{19} - \vec{e}_{20} \nabla_{20} - \vec{e}_{21} \nabla_{21}$$
$$- \vec{e}_{22} \nabla_{22} - \vec{e}_{23} \nabla_{23} - \vec{e}_{24} \nabla_{24} - \vec{e}_{25} \nabla_{25} - \vec{e}_{26} \nabla_{26}$$
$$- \vec{e}_{27} \nabla_{27} - \vec{e}_{28} \nabla_{28} - \vec{e}_{29} \nabla_{29} - \vec{e}_{30} \nabla_{30} - \vec{e}_{31} \nabla_{31} \tag{38}$$

## 5.1 Field potential and strength

In the trigintaduonion field T-B, the field potential $\mathcal{A} = (a_0, a_1, a_2, a_3, a_4, a_5, a_6, a_7, a_8, a_9, a_{10}, a_{11}, a_{12}, a_{13}, a_{14}, a_{15}, a_{16}, a_{17}, a_{18}, a_{19}, a_{20}, a_{21}, a_{22}, a_{23}, a_{24}, a_{25}, a_{26}, a_{27}, a_{28}, a_{29}, a_{30}, a_{31})$ is defined as ($\diamondsuit = \diamondsuit_{T-B}$)

$$\mathcal{A} = \diamondsuit^* \circ \mathcal{X}$$
$$= a_0 + a_1 \vec{e}_1 + a_2 \vec{e}_2 + a_3 \vec{e}_3 + a_4 \vec{e}_4 + a_5 \vec{e}_5 + a_6 \vec{e}_6$$
$$+ a_7 \vec{e}_7 + a_8 \vec{e}_8 + a_9 \vec{e}_9 + a_{10} \vec{e}_{10} + a_{11} \vec{e}_{11}$$
$$+ a_{12} \vec{e}_{12} + a_{13} \vec{e}_{13} + a_{14} \vec{e}_{14} + a_{15} \vec{e}_{15} + a_{16} \vec{e}_{16}$$
$$+ a_{17} \vec{e}_{17} + a_{18} \vec{e}_{18} + a_{19} \vec{e}_{19} + a_{20} \vec{e}_{20} + a_{21} \vec{e}_{21}$$
$$+ a_{22} \vec{e}_{22} + a_{23} \vec{e}_{23} + a_{24} \vec{e}_{24} + a_{25} \vec{e}_{25} + a_{26} \vec{e}_{26}$$
$$+ a_{27} \vec{e}_{27} + a_{28} \vec{e}_{28} + a_{29} \vec{e}_{29} + a_{30} \vec{e}_{30} + a_{31} \vec{e}_{31} \tag{39}$$

where, the mark (*) denotes the trigintaduonion conjugate.

The field strength $\mathcal{B}$ of the trigintaduonion field T-B can be defined as

$$\mathcal{B} = \diamondsuit \circ \mathcal{A}$$
$$= (\nabla_0 + \vec{e}_1 \nabla_1 + \vec{e}_2 \nabla_2 + \vec{e}_3 \nabla_3 + \vec{e}_4 \nabla_4 + \vec{e}_5 \nabla_5$$
$$+ \vec{e}_6 \nabla_6 + \vec{e}_7 \nabla_7 + \vec{e}_8 \nabla_8 + \vec{e}_9 \nabla_9 + \vec{e}_{10} \nabla_{10} + \vec{e}_{11} \nabla_{11}$$
$$+ \vec{e}_{12} \nabla_{12} + \vec{e}_{13} \nabla_{13} + \vec{e}_{14} \nabla_{14} + \vec{e}_{15} \nabla_{15} + \vec{e}_{16} \nabla_{16}$$
$$+ \vec{e}_{17} \nabla_{17} + \vec{e}_{18} \nabla_{18} + \vec{e}_{19} \nabla_{19} + \vec{e}_{20} \nabla_{20} + \vec{e}_{21} \nabla_{21}$$
$$+ \vec{e}_{22} \nabla_{22} + \vec{e}_{23} \nabla_{23} + \vec{e}_{24} \nabla_{24} + \vec{e}_{25} \nabla_{25} + \vec{e}_{26} \nabla_{26}$$
$$+ \vec{e}_{27} \nabla_{27} + \vec{e}_{28} \nabla_{28} + \vec{e}_{29} \nabla_{29} + \vec{e}_{30} \nabla_{30} + \vec{e}_{31} \nabla_{31})$$
$$\circ (a_0 + a_1 \vec{e}_1 + a_2 \vec{e}_2 + a_3 \vec{e}_3 + a_4 \vec{e}_4 + a_5 \vec{e}_5$$
$$+ a_6 \vec{e}_6 + a_7 \vec{e}_7 + a_8 \vec{e}_8 + a_9 \vec{e}_9 + a_{10} \vec{e}_{10} + a_{11} \vec{e}_{11}$$
$$+ a_{12} \vec{e}_{12} + a_{13} \vec{e}_{13} + a_{14} \vec{e}_{14} + a_{15} \vec{e}_{15} + a_{16} \vec{e}_{16}$$
$$+ a_{17} \vec{e}_{17} + a_{18} \vec{e}_{18} + a_{19} \vec{e}_{19} + a_{20} \vec{e}_{20} + a_{21} \vec{e}_{21}$$
$$+ a_{22} \vec{e}_{22} + a_{23} \vec{e}_{23} + a_{24} \vec{e}_{24} + a_{25} \vec{e}_{25} + a_{26} \vec{e}_{26}$$
$$+ a_{27} \vec{e}_{27} + a_{28} \vec{e}_{28} + a_{29} \vec{e}_{29} + a_{30} \vec{e}_{30} + a_{31} \vec{e}_{31}) \tag{40}$$

## 5.2 Field source

The field source of the trigintaduonion field T-B can be defined as

$$\mu \mathcal{S} = (\mathcal{B}/\mathbf{K} + \diamondsuit)^* \circ \mathcal{B}$$
$$= (\mathcal{B}_{H-S} / k^B_{H-S} + \mathcal{B}_{S-W} / k^B_{S-W} + \mathcal{B}_{E-G} / k^B_{E-G} + \mathcal{B}_{H-W} / k^B_{H-W} + \diamondsuit)^* \circ \mathcal{B}$$
$$= (\diamondsuit^* \circ \diamondsuit) \circ \mathcal{A}$$
$$+ (\mathcal{B}_{H-S} / k^B_{H-S} + \mathcal{B}_{S-W} / k^B_{S-W} + \mathcal{B}_{E-G} / k^B_{E-G} + \mathcal{B}_{H-W} / k^B_{H-W})^* \circ \mathcal{B} \tag{41}$$

where, $\mathbf{K} = \mathbf{K}_{T-B}$, $k^B_{H-S}$, $k^B_{S-W}$, $k^B_{E-G}$ and $k^B_{H-W}$ are coefficients in the trigintaduonion space. The coefficient $\mu$ is interaction intensity of trigintaduonion field T-B. The field strengths are

$$\mathcal{B}_{E-G} = b_0 + b_1 \vec{e}_1 + b_2 \vec{e}_2 + b_3 \vec{e}_3 + b_4 \vec{e}_4 + b_5 \vec{e}_5 + b_6 \vec{e}_6 + b_7 \vec{e}_7$$
$$\mathcal{B}_{S-W} = b_8 \vec{e}_8 + b_9 \vec{e}_9 + b_{10} \vec{e}_{10} + b_{11} \vec{e}_{11} + b_{12} \vec{e}_{12} + b_{13} \vec{e}_{13} + b_{14} \vec{e}_{14} + b_{15} \vec{e}_{15}$$
$$\mathcal{B}_{H-S} = b_{16} \vec{e}_{16} + b_{17} \vec{e}_{17} + b_{18} \vec{e}_{18} + b_{19} \vec{e}_{19} + b_{20} \vec{e}_{20} + b_{21} \vec{e}_{21} + b_{22} \vec{e}_{22} + b_{23} \vec{e}_{23}$$
$$\mathcal{B}_{H-W} = b_{24} \vec{e}_{24} + b_{25} \vec{e}_{25} + b_{26} \vec{e}_{26} + b_{27} \vec{e}_{27} + b_{28} \vec{e}_{28} + b_{29} \vec{e}_{29} + b_{30} \vec{e}_{30} + b_{31} \vec{e}_{31}$$

In the above Eq.(41),

$$(\nabla_m \nabla_k - \nabla_k \nabla_m) a^i = \Sigma R_{kmp}{}^i a^p \tag{42}$$
$$R_{kmp}{}^i = \partial_m \Gamma_{pk}{}^i - \partial_k \Gamma_{pm}{}^i + \Sigma \Gamma_{qm}{}^i \Gamma_{pk}{}^q - \Sigma \Gamma_{qk}{}^i \Gamma_{pm}{}^q \tag{43}$$

wherein, $R_{kmp}{}^i$ is the curvature of curved trigintaduonion space; $m, n, p, q = 0, 1, 2, \ldots, 31$.

From the above equation, we can obtain the field equations $R_{kmp}{}^i = f(k^B{}_{H-S}, k^B{}_{S-W}, k^B{}_{E-G}, k^B{}_{H-W}, \mu S, \mathcal{B}, \mathcal{A}, K)$. In the curved trigintaduonion space ($\diamondsuit = \diamondsuit_{T-B}$), other physical quantities in the curved trigintaduonion field T-B including the coefficient $\Gamma_{pk}{}^i$ and the curvature $R_{kmp}{}^i$ can be obtained in Table 13. Where, the coefficient $\mathcal{B}/K = \mathcal{B}_{E-G}/k^B{}_{E-G} + \mathcal{B}_{S-W}/k^B{}_{S-W} + \mathcal{B}_{H-S}/k^B{}_{H-S} + \mathcal{B}_{H-W}/k^B{}_{H-W}$, and coefficient $\mathcal{W}/K\,b = \mathcal{W}_{E-G}/k^B{}_{E-G}\,b^B{}_{E-G} + \mathcal{W}_{S-W}/k^B{}_{S-W}\,b^B{}_{S-W} + \mathcal{W}_{H-S}/k^B{}_{H-S}\,b^B{}_{H-S} + \mathcal{W}_{H-W}/k^B{}_{H-W}\,b^B{}_{H-W}$.

Table 13. The curvature of curved trigintaduonion space with other physical quantities

|   | physical quantity | coefficient, $\Gamma_{pk}{}^i$ | curvature, $R_{kmp}{}^i$ |
|---|---|---|---|
| 1 | $\mathcal{X}$ | $\mathcal{A} = \diamondsuit^* \circ \mathcal{X}$ | $\mathcal{B} = \diamondsuit^2 \circ \mathcal{X}$ |
| 2 | $\mathcal{A}$ | $\mathcal{B} = \diamondsuit \circ \mathcal{A}$ | $\mu S = \diamondsuit^2 \circ \mathcal{A} + (\mathcal{B}/K)^* \circ \mathcal{B}$ |
| 3 | $\mathcal{B}$ | $\mu S = (\diamondsuit + \mathcal{B}/K)^* \circ \mathcal{B}$ | $\mu Z = K(\diamondsuit + \mathcal{B}/K)^2 \circ \mathcal{B}$ |
| 4 | $\mathcal{M}$ | $\mathcal{W} = K(\diamondsuit + \mathcal{B}/K)^* \circ \mathcal{M}$ | $\mathcal{N} = K^2(\diamondsuit + \mathcal{B}/K)^2 \circ \mathcal{M}$ |
| 5 | $\mathcal{B}$ | $\mathcal{T} = (\diamondsuit + \mathcal{W}/K\,b)^* \circ \mathcal{B}$ | $O = (\diamondsuit + \mathcal{W}/K\,b)^2 \circ \mathcal{B}$ |
| 6 | $\mathcal{M}$ | $\mathcal{U} = (\diamondsuit + \mathcal{W}/K\,b)^* \circ \mathcal{M}$ | $\mathcal{L} = (\diamondsuit + \mathcal{W}/K\,b)^2 \circ \mathcal{M}$ |

5.3 Other fields in the curved trigintaduonion space

With the similar method, we can obtain the field equations of the trigintaduonion field T-X, the trigintaduonion field T-A and trigintaduonion field T-S in the local curved trigintaduonion space respectively.

Table 14. The curvature and other physical quantities of trigintaduonion field T-X

|   | physical quantity | coefficient, $\Gamma_{pk}{}^i$ | curvature, $R_{kmp}{}^i$ |
|---|---|---|---|
| 1 | $\mathcal{X}$ | $\mathcal{A} = (\diamondsuit + \mathcal{X}/K)^* \circ \mathcal{X}$ | $\mathcal{B} = (\diamondsuit + \mathcal{X}/K)^2 \circ \mathcal{X}$ |
| 2 | $\mathcal{A}$ | $\mathcal{B} = (\diamondsuit + \mathcal{X}/K) \circ \mathcal{A}$ | $\mu S = (\diamondsuit + \mathcal{X}/K)^2 \circ \mathcal{A}$ |
| 3 | $\mathcal{B}$ | $\mu S = (\diamondsuit + \mathcal{X}/K)^* \circ \mathcal{B}$ | $\mu Z = K(\diamondsuit + \mathcal{X}/K)^2 \circ \mathcal{B}$ |
| 4 | $\mathcal{M}$ | $\mathcal{W} = K(\diamondsuit + \mathcal{X}/K)^* \circ \mathcal{M}$ | $\mathcal{N} = K(\diamondsuit + \mathcal{X}/K)^2 \circ \mathcal{M}$ |
| 5 | $\mathcal{X}$ | $\mathcal{D} = (\diamondsuit + \mathcal{W}/K\,b)^* \circ \mathcal{X}$ | $\mathcal{G} = (\diamondsuit + \mathcal{W}/K\,b)^2 \circ \mathcal{X}$ |
| 6 | $\mathcal{D}$ | $\mathcal{G} = (\diamondsuit + \mathcal{W}/K\,b) \circ \mathcal{D}$ | $\mathcal{T} = (\diamondsuit + \mathcal{W}/K\,b)^2 \circ \mathcal{D}$ |
| 7 | $\mathcal{G}$ | $\mathcal{T} = (\diamondsuit + \mathcal{W}/K\,b)^* \circ \mathcal{G}$ | $O = (\diamondsuit + \mathcal{W}/K\,b)^2 \circ \mathcal{G}$ |
| 8 | $\mathcal{M}$ | $\mathcal{U} = (\diamondsuit + \mathcal{W}/K\,b)^* \circ \mathcal{M}$ | $\mathcal{L} = (\diamondsuit + \mathcal{W}/K\,b)^2 \circ \mathcal{M}$ |

In the curved trigintaduonion space, other physical quantities in the curved trigintaduonion field T-X including coefficient $\Gamma_{pk}{}^i$ and/or curvature $R_{kmp}{}^i$ can be obtained in the Table 14. Wherein, the coefficient $\mathcal{X}/K = \mathcal{X}_{E-G}/k^X{}_{E-G} + \mathcal{X}_{S-W}/k^X{}_{S-W} + \mathcal{X}_{H-S}/k^X{}_{H-S} + \mathcal{X}_{H-W}/k^X{}_{H-W}$, and the coefficient $\mathcal{W}/Kb = \mathcal{W}_{E-G}/k^X{}_{E-G}b^X{}_{E-G} + \mathcal{W}_{S-W}/k^X{}_{S-W}b^X{}_{S-W} + \mathcal{W}_{H-S}/k^X{}_{H-S}b^X{}_{H-S} + \mathcal{W}_{H-W}/k^X{}_{H-W}b^X{}_{H-W}$.

In the curved trigintaduonion space, other physical quantities in the curved trigintaduonion field T-A including coefficient $\Gamma_{pk}{}^i$ and/or curvature $R_{kmp}{}^i$ can be obtained in the Table 15. Wherein, the coefficient $\mathcal{A}/K = \mathcal{A}_{E-G}/k^A{}_{E-G} + \mathcal{A}_{S-W}/k^A{}_{S-W} + \mathcal{A}_{H-S}/k^A{}_{H-S} + \mathcal{A}_{H-W}/k^A{}_{H-W}$, and the coefficient $\mathcal{W}/Kb = \mathcal{W}_{E-G}/k^A{}_{E-G}b^A{}_{E-G} + \mathcal{W}_{S-W}/k^A{}_{S-W}b^A{}_{S-W} + \mathcal{W}_{H-S}/k^A{}_{H-S}b^A{}_{H-S} + \mathcal{W}_{H-W}/k^A{}_{H-W}b^A{}_{H-W}$.

Table 15.   The curvature and other physical quantities of trigintaduonion field T-A

|   | physical quantity | coefficient, $\Gamma_{pk}{}^i$ | curvature, $R_{kmp}{}^i$ |
|---|---|---|---|
| 1 | $\mathcal{X}$ | $\mathcal{A} = \diamondsuit^* \circ \mathcal{X}$ | $\mathcal{B} = \diamondsuit^2 \circ \mathcal{X}$ |
| 2 | $\mathcal{A}$ | $\mathcal{B} = (\diamondsuit + \mathcal{A}/K) \circ \mathcal{A}$ | $\mu\mathcal{S} = (\diamondsuit + \mathcal{A}/K)^2 \circ \mathcal{A}$ |
| 3 | $\mathcal{B}$ | $\mu\mathcal{S} = (\diamondsuit + \mathcal{A}/K)^* \circ \mathcal{B}$ | $\mu\mathcal{Z} = K(\diamondsuit + \mathcal{A}/K)^2 \circ \mathcal{B}$ |
| 4 | $\mathcal{M}$ | $\mathcal{W} = K(\diamondsuit + \mathcal{A}/K)^* \circ \mathcal{M}$ | $\mathcal{N} = K(\diamondsuit + \mathcal{A}/K)^2 \circ \mathcal{M}$ |
| 5 | $\mathcal{A}$ | $\mathcal{G} = (\diamondsuit + \mathcal{W}/Kb) \circ \mathcal{A}$ | $\mathcal{T} = (\diamondsuit + \mathcal{W}/Kb)^2 \circ \mathcal{A}$ |
| 6 | $\mathcal{G}$ | $\mathcal{T} = (\diamondsuit + \mathcal{W}/Kb)^* \circ \mathcal{G}$ | $O = (\diamondsuit + \mathcal{W}/Kb)^2 \circ \mathcal{G}$ |
| 7 | $\mathcal{M}$ | $\mathcal{U} = (\diamondsuit + \mathcal{W}/Kb)^* \circ \mathcal{M}$ | $\mathcal{L} = (\diamondsuit + \mathcal{W}/Kb)^2 \circ \mathcal{M}$ |

In the curved trigintaduonion space, other physical quantities in curved trigintaduonion field T-S including coefficient $\Gamma_{pk}{}^i$ and/or curvature $R_{kmp}{}^i$ can be obtained in Table 16. Where, the coefficient $\mathcal{S}/K = \mathcal{S}_{E-G}/k^S{}_{E-G} + \mathcal{S}_{S-W}/k^S{}_{S-W} + \mathcal{S}_{H-S}/k^S{}_{H-S} + \mathcal{S}_{H-W}/k^S{}_{H-W}$, and $\mathcal{W}/Kb = \mathcal{W}_{E-G}/k^S{}_{E-G}b^S{}_{E-G} + \mathcal{W}_{S-W}/k^S{}_{S-W}b^S{}_{S-W} + \mathcal{W}_{H-S}/k^S{}_{H-S}b^S{}_{H-S} + \mathcal{W}_{H-W}/k^S{}_{H-W}b^S{}_{H-W}$.

Table 16.   The curvature and other physical quantities of trigintaduonion field T-S

|   | physical quantity | coefficient, $\Gamma_{pk}{}^i$ | curvature, $R_{kmp}{}^i$ |
|---|---|---|---|
| 1 | $\mathcal{X}$ | $\mathcal{A} = \diamondsuit^* \circ \mathcal{X}$ | $\mathcal{B} = \diamondsuit^2 \circ \mathcal{X}$ |
| 2 | $\mathcal{A}$ | $\mathcal{B} = \diamondsuit \circ \mathcal{A}$ | $\mu\mathcal{S} = \diamondsuit^2 \circ \mathcal{A}$ |
| 3 | $\mathcal{B}$ | $\mu\mathcal{S} = \diamondsuit^* \circ \mathcal{B}$ | $\mu\mathcal{Z} = K \diamondsuit^2 \circ \mathcal{B} + \mu K(\mathcal{S}/K) \circ \mathcal{S}$ |
| 4 | $\mathcal{M}$ | $\mathcal{W} = K(\diamondsuit + \mathcal{S}/K)^* \circ \mathcal{M}$ | $\mathcal{N} = K(\diamondsuit + \mathcal{S}/K)^2 \circ \mathcal{M}$ |
| 5 | $\mathcal{M}$ | $\mathcal{U} = (\diamondsuit + \mathcal{W}/Kb)^* \circ \mathcal{M}$ | $\mathcal{L} = (\diamondsuit + \mathcal{W}/Kb)^2 \circ \mathcal{M}$ |

## 6. Conclusions

The paper describes the field theories and quantum theories in the quaternion, octonion, sedenion and trigintaduonion spacetimes, including the coefficients $\Gamma_{pk}{}^i$ and curvature $R_{kmp}{}^i$ in different kinds of the curved spacetimes.

In the field theories of the curved octonion spacetimes etc, we extend the concept of curved spacetime from the Four-spacetime to quaternion space etc, and deduce some conclusions which are consistent with Einsteinian General Relativity.

In the quantum theories of the curved octonion spacetime etc, we expand the quantum theory from the flat spacetime to the curved spacetime, and obtain some predictions of quantum theories in the octonion spacetime etc.

**Acknowledgements**

The author thanks Xu Chen, Shaohan Lin, Minfeng Wang, Yun Zhu and Zhimin Chen for helpful discussions. This project was supported by National Natural Science Foundation of China under grant number 60677039, Science & Technology Department of Fujian Province of China under grant number 2005HZ1020 and 2006H0092, and Xiamen Science & Technology Bureau of China under grant number 3502Z20055011.

**References**

[1] Zihua Weng. Octonionic electromagnetic and gravitational interactions and dark matter. arXiv: physics /0612102.
[2] Zihua Weng. Octonionic quantum interplays of dark matter and ordinary matter. arXiv: physics /0702019.
[3] Luisa T. Buchman, Olivier C. A. Sarbach. Towards absorbing outer boundaries in General Relativity. Class. Quant. Grav., 23 (2006) 6709-6744.
[4] Zihua Weng. Octonionic hyper-strong and hyper-weak fields and their quantum equations. arXiv: physics /0702086.
[5] Nicolas Boulanger, Fabien Buisseret, Philippe Spindel. On bound states of Dirac particles in gravitational fields. Phys. Rev. D, 74 (2006) 125014.
[6] Zihua Weng. Compounding fields and their quantum equations in the sedenion space. arXiv: physics /0703055.
[7] Zihua Weng. Compounding Fields and Their Quantum Equations in the Trigintaduonion Space. arXiv: 0704.0136.